\documentclass{article}
\usepackage{graphicx}
\usepackage{psfrag}
\usepackage{bm}
\usepackage{color} 
\usepackage{amsmath}
\usepackage{amssymb}
\usepackage{amsfonts}
\usepackage{float}

\title{Dean flow and vortex shedding in a three-dimensional 180$^\circ$ sharp bend}
\author{Alban Poth\'erat$^{1}$
and  Lintao Zhang$^{2}$\\
$^1$
Applied Mathematics Research Centre, Coventry University,\\ Priory Street, Coventry, CV1 5FB, United Kingdom\\
$^2$
Advanced Sustainable Manufacturing Technologies (ASTUTE 2020),\\ 
College of Engineering, Swansea University,\\
 Bay Campus, Fabian Way, Swansea SA1 8EN, UK\\
}

\begin{document}

\maketitle

\begin{abstract}
We present a detailed analysis of the flow in a 180$^o$ sharp bend of square cross-section. Besides 
numerous applications where this generic configuration is found, its main fundamental interest resides
in the co-existence of a recirculation bubble in the outlet and a pair of Dean vortices driven from 
within the turning part of the bend, and how their interaction may drive the flow dynamics.
A critical point analysis first shows that the complex flow topology that results from this 
particular configuration can be reduced to three pairs of critical points in the symmetry plane of the 
bend (with a focus and a half-saddle each). These pairs respectively generate the first recirculation 
bubble, the pair of Dean vortex tubes and a third pair of vortex tubes located in the upper corner of the bend, akin to the Dean vortices but of much lower intensity and impact on the rest of the flow.\\
The Dean flow by contrast drives a strong vertical jet that splits the recirculation bubble into 
two symmetric lobes. Unsteadiness sets in at $Re\lesssim800$ through a supercritical bifurcation, as these lobes start oscillating antisymmetrically. These initially periodic oscillations grow 
in amplitude until the lobes break away from the main recirculation. The system then settles into 
a second periodic state where streamwise vortices driven by the Dean flow are alternatively formed and shed 
on the left and right part of the outlet. This novel mechanism of periodic vortex shedding results 
 from the subtle interaction between the recirculation bubble in the outlet and the pair of Dean vortices 
generated in the turning part, and in this sense, they are expected to be a unique feature of the 180$^o$ bend with sufficiently close side walls. 

\end{abstract}

\section{Introduction \label{sec:intro}}
This study focuses on the flow in a $180^o$ sharp bend of square cross-section,
 in regimes where the flow is either steady or slightly beyond the onset of 
unsteadiness. 
The interest in sharp bends chiefly arises from the optimisation of heat 
exchangers (\cite{chung2003_ijhff}), whose thermal efficiency is driven by the 
internal flow structure.
Applications involve a great variety of features, flow regimes, 
and dimensions ranging from microfluidics (\cite{huang2014_jfm}) to the cooling 
of nuclear fusion reactors where magnetic fields can modify the flow 
(\cite{mistrangelo2011_fed,shps2014_afmc}). The flow structure results from the 
combined influence of the two main features of the problem. First, flow 
separation occurs even a low Reynolds number near the sharp inner corner of the bend and leads to the 
at least one recirculating region in the bend outlet (\cite{zp2013_pf}). Second, 
the centrifulgal force 
in the turning part 
drives secondary flows, first identified by \cite{dean1927_pm} in curve bends,
that return across the entire bend section 
(\cite{berger1983_arfm}). 
Because of the 
inherent complexity that ensues, two partial approaches have been preferred to 
a systematic analysis of the full problem until now.
In the first approach, recirculating and secondary flows were analysed 
separately but systematically. In the second, single aspects of the full 
problem, mainly linked to heat transfer, have been tackled in view of particular applications.\\
The first approach is more general because of the generic nature of 
recirculating regions behind flow separations, and of secondary flows. 
The former also occur in the wake of obstacles (\cite{williamson1996_arfm}) and 
behind backward-facing steps (BFS) (\cite{armaly1983_jfm}). In ideal 
configurations 
without side walls, the length of the recirculating bubble increases 
practically linearly with the Reynolds number in the steady regime 
(\cite{williamson1996_arfm}) and collapses 
at the onset of unsteadiness (see \cite{chung2003_ijhff} for $180^o$ bends). 
If a second wall is present opposite the first bubble (in sharp bends and BFS), flow 
expansion behind the bubble promotes a second region with a flow 
separation and a recirculation on the opposite wall  
(\cite{barkley2002_jfm,zp2013_pf}). 
When the Reynolds number exceeds a critical value, instabilities occur in the region of the 
first bubble that trigger a transition to unsteadiness in all three configurations. The 
conditions of this transition are however very sensitive to the geometry. 
Behind unconfined cylinders, the periodic vortex shedding organised in a von K\`arman 
street appears at $Re=46$,
but in sharp bends and BFS, both the critical Reynolds number and the nature of 
the critical mode heavily depend on the \emph{opening ratio}, $\beta$  between 
the minimum and maximum channel width. At low values of $\beta$, a jet-like instability with oscillatory critical modes takes place, whereas for $\beta$ near unity and beyond, the instable mode is localised within the bubble itself, with no oscillatory component 
(\cite{lanzerstorfer2012_jfm, shps2016_jfm}). Crucially, these results were 
obtained in configurations without side walls, for which the instability 
develops on a two-dimensional base state. In these cases, three-dimensionality 
appears only in the unstable modes (\cite{kaikstis1991_jfm,shps2016_jfm}). In bends of rectangular 
cross-section,  by contrast the base flow itself must be three-dimensional 
to satisfy the no-slip condition at the side walls and the mechanisms of transition 
to unsteadiness are unknown.\\
Similarly, the second major feature present in sharp bends of square 
cross-section (the secondary flows) have been extensively studied since 
Dean's original work showing the existence of counter-rotating vortices of 
streamwise rotating axis in flow near curved boundaries 
(\cite{dean1927_pm,dean1928_pm}). Their occurrence has been mostly studied in 
smooth rather than sharp bends, where flow separation is mostly absent. A 
comprehensive review on the topic can be found in \cite{berger1983_arfm}. In 
the context of the sharp bend, an interesting feature of secondary 
flows is that Dean vortices (DV) are a very robust: although their exact shape does depend on geometry and Reynolds number, they 
have been observed in a ducts of a great variety of cross sections, in laminar  
as well as highly turbulent regimes (For square sections and various bend curvatures, see analytical solutions by  \cite{ito1951_rihsm,cuming1952_arc} and 
experiments by \cite{schabacker1998_asme} at $Re=5\times10^4$). Also, 
\cite{joseph1975_aichej} found an interesting structure with four Dean 
vortices at 
moderate Reynolds numbers, still in a curved bend of square cross-section. This 
stresses that the most famous picture of two-counter-rotating Dean vortices is 
by no means the only possible topology in curved geometries.\\ 
The great challenge of the sharp bend is to understand how these 
remarkable, and well understood features of their own interact.  A number of 
studies tackled the full problem of sharp bends of square cross sections, 
mainly in view of characterising their heat transfer properties. These bring 
some indications, but no definite answers to this question. First, 
both main recirculation and the Dean flow do coexist over wide range of 
laminar and turbulent regimes (\cite{mochizuki1999_ijhmt}). Nevertheless, the
main recirculation appears distorted by the presence of the Dean flow, which 
raises further questions on the mechanism governing the transition to 
unsteadiness.
Consequently, very basic questions remain open regarding the topology and the dynamics of flows in sharp bends:
\begin{enumerate}
\item What is the precise topology of the flow when a main recirculation coexists with secondary flows ?
\item What are the conditions in which both these structures co-exist ?
\item Which mechanism underpins the onset of unsteadiness ?
\item How does the flow dynamics translate in terms of global, measurable quantities such as drag/lift coefficients on the bend elements, and Strouhal number 
(measuring the main flow frequencies) ?
\end{enumerate}
%
%
We tackle these questions by means of a parametric study based on  Direct 
Numerical Simulations of the flow in a duct of opening ratio $\beta=1$, 
increasing the Reynolds number from 5 to 2000. 
After a brief description of the problem and 
validation of the numerical methods in section \ref{sec:cn}, 
we determine the main structures governing the dynamics of the steady flow 
regimes by means of a systematic analysis of the topology based on its  critical points (section \ref{sec:steady}). We then characterise the onset of unsteadiness and the bifurcation 
leading to it, using a simple model based on the Landau equation following the 
ideas of \cite{sheard2004_jfm} (section \ref{sec:unsteady}). 
Finally, we examine how the succession of regimes affects flow coefficients (lift, and drag coefficients on the separating wall, Strouhal numbers), in 
view of offering a simple way of detecting their occurrence in practical 
situations (section \ref{sec:coefs}).

\section{Configuration and numerical set-up}
\label{sec:cn}
\subsection{Configuration}
\begin{figure}
\psfrag{a}{$a$}
\psfrag{b}{$b$}
\psfrag{c}{$c$}
\psfrag{d}{$d$}
\psfrag{ee}{$e$}
\psfrag{ff}{$f$}
\psfrag{oo}{$O$}
\psfrag{x}{$x$}
\psfrag{y}{$y$}
\psfrag{z}{$z$}
\psfrag{Inlet}{Inlet}
\psfrag{Outlet}{Outlet}
\psfrag{End wall}{End wall}
\begin{center}
\includegraphics[width=4in]{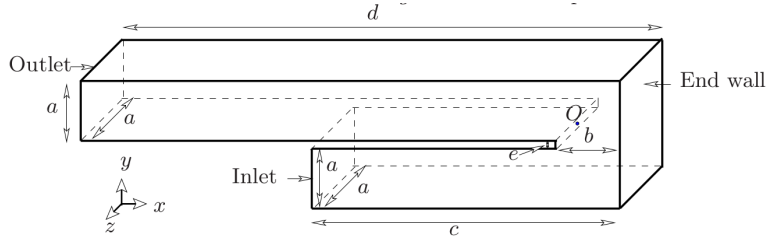}
\end{center}
\caption{The geometry of three-dimensional $180^\circ$ sharp bend: $b=a$, $c=10a$, $d=30a$ and $e=0.04a$.}
\label{fig:sketch}
\end{figure}

We consider an incompressible flow (fluid density $\rho$, kinematic viscosity 
$\nu$) in a 3D $180^\circ$ sharp bend of square cross section of size $a$, 
represented in figure \ref{fig:sketch}.
The origin of the frame of reference $O$, is taken at the inside centre of the 
turning part.
The lengths of the turning part, inlet and outlet are $b$, $c-b$, and $d-b$. 
The divider thickness is $e$. In the present paper, a fixed geometry is considered, with an opening ratio $b/a$ set to 1, $c=10a$, $d=30a$ and $e=0.04a$. 
The choice of $c=10a$ ensures that the flow reaches a fully developed state in the inlet, regardless of the choice of inlet profile and for the range of Reynolds numbers considered in this paper (This was verified with simulations at $Re$=100 and $Re=500$). Following \cite{sohankar1998_ijnmhff}'s recommendation for cylinder wakes, 
$d=30a$ is chosen so that all vortical structures shed in the turning part have 
been damped out before the flow reaches the outlet, as we did previously in 
\cite{zp2013_pf}. With a divider thickness of $e=0.04a$, the bend is sharp, and 
the exact value of this parameter can be expected to have negligible influence on the flow features.
\subsection{Flow equations and numerical set-up}
\begin{figure}
\psfrag{i}{Inlet}
\psfrag{o}{Outlet}
\psfrag{e}{End wall}
\psfrag{x}{$x$}
\psfrag{y}{$y$}
\begin{center}
\includegraphics[width=3in]{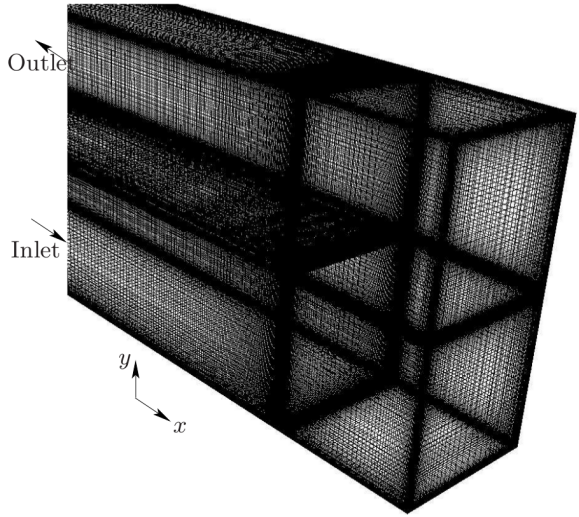}
\end{center}
\caption{General view of mesh M0.}
\label{fig:mesh}
\end{figure}

The flow is  governed by the Navier-Stokes equations, which are written in 
non-dimensional form as:
\begin{equation}
\label{equ:NSe} 
\partial_t{\mathbf{u}}+({\mathbf{u}} \cdotp \nabla ){\mathbf{u}}+\nabla p  = \frac{1}{Re} \nabla^2{\mathbf{u}},
\end{equation}
\begin{equation}
\label{equ:NSe2}
\nabla\cdotp\textbf{u}=0,
\end{equation}
where $\mathbf{u}$ and $p$ are the non-dimensional flow velocity and pressure, built using the maximum inlet velocity $U_0$ as reference velocity, and $a$ as  reference length. $Re$=$U_{0}a/\nu$ is the Reynolds number. A no-slip impermeable boundary condition is imposed at all solid walls through a homogeneous Dirichlet condition for the velocity. A homogeneous Neumann condition for the velocity is applied for the velocity at the outlet. A three dimensional Poiseuille velocity profile is imposed at the inlet:
\begin{equation}
\textbf{u}_{x}(y,z)=U_0[1-(\frac{2(y+0.052)}{a})^2][1-(\frac{2(z+0.05)}{a})^2].
\end{equation}
Though easy to implement, this inlet condition is not an exact solution of the fully established flow in a straight duct. Nevertheless, the flow is always regularised before it hits the turning part (see \cite{fmwhite2005} p. 120).
This approach saves the preliminary calculation that would have 
been required to establish the numerical solution in a straight duct for each 
simulated case.\\
We investigate this problem by means of three-dimensional direct numerical 
simulations with a finite-volume code based on the OpenFOAM 1.6 framework. The 
code is detailed and has been thoroughly validated by \cite{dp10_jfm} for a range 
of Reynolds numbers comparable to the ones we investigate here. The flow 
equations (\ref{equ:NSe}) and (\ref{equ:NSe2}) are solved in a segregated way 
and the PISO algorithm detailed in \cite{weller98} is adopted to deal with 
the pressure-velocity coupling. For the pressure boundary condition, a 
homogeneous Neumann condition is imposed at all boundaries but the outlet, 
where a Dirichlet condition is applied. During the simulations, the time step 
was kept constant, so as to satisfy the Courant-Lewy-Friedrich condition, such 
that the maximum Courant number is always smaller than 1. The mesh, shown in figure \ref{fig:mesh}, is 
fully structured and is refined in the vicinity of the walls (down to a cell size of $0.0035a$ and 
$0.002a$ with a ratio of 0.025 and 0.05 between wall and centre cells, in geometric progression over 60,
320, 180 in the $z$, $x$ and $y$ directions, respectively.).
The mesh was validated against finer meshes for which resolution was separately increased in all three directions of space. 
The main characteristics of the tested meshes are provided in table \ref{tab:nodeselect}. 
\begin{table}
\begin{center}
\begin{tabular}{lcccc}
Meshes                 & Mesh 1  & Mesh 2 & Mesh 3& Mesh 0 \\[3pt]
Total number of nodes  & 6125681  & 6104941  & 6086421 &3068361\\
$\epsilon_{C_{d}}=|1-C_{d}(M_1)/C_{d}(M_{i})| $ & 8.7 $\times$ 10$^{-4}$ & 7.7 $\times$ 10$^{-4}$ & 7.4 $\times$ 10$^{-4}$ &-\\
$\epsilon_{C_{st}}=|1-St(Mi)/St(M_{0})| $  & 3.5 $\times$ 10$^{-2}$ & 3.5 $\times$ 10$^{-2}$  & 2.7 $\times$ 10$^{-2}$ &-\\  
\end{tabular}      
\caption{Main characteristics of the different meshes and errors in drag coefficient $C_d$ and Strouhal number $St$ relative to the reference mesh M0 at $Re$=1000.M1, M2 and M3 have double resolution to RM along $\textbf{e}_x$, $\textbf{e}_y$ and $\textbf{e}_z$, respectively.}
\label{tab:nodeselect}
\end{center}
\end{table}
The results show that Mesh 1 provides good accuracy, whilst 
keeping computational costs reasonable enough for a parametric analysis on the 
Reynolds number.\\

We carried out several successive simulations at increasing $Re$ in the range [5-2000]. In each case, the initial conditions were taken from either the steady state, or from a snapshot of the fully developed unsteady state obtained at the previous value of $Re$. For unsteady cases, the flows were computed over a total simulation time of around 100 shedding times. Our computation yielded steady flow regimes for $Re\leq700$ and unsteady flow regimes for $Re \ge 800$.\\
\subsection{Analysis of the flow topology}
In order to extricate the complex topology of the flow structures, we shall 
rely on the critical point analysis introduced by \cite{hunt1978_jfm}. The main 
idea is to seek critical points of the stress field at no-slip walls and 
critical
points of the velocity field in symmetry planes, where streamlines separate. 
This way, critical points naturally distinguish groups of streamlines forming
distinct flow structures. We found that all critical points of interest for the main dynamics of the flow were captured by 
considering streamlines originating in the inlet region of the vertical 
centreplane $(x,y,0)$ (CP) or converging to the outlet region of the same plane, 
as for the flow around a confined obstacle (\cite{dp10_jfm,dp12_jfm}). Hence we shall 
focus our analysis on critical points in the CP. Critical points in the stresslines along walls shall not be systematically discussed.\\

To identify the recirculation and vortex structures, we shall use the classical approach based on 
 the eigenvalues of the symmetric tensor ${\bf T}={\bf S}^2+ {\bf \Omega}^2$, where
$\bf S$ and $\bf \Omega$ are respectively the symmetric and antisymmetric part of 
the velocity gradient tensor $\nabla\mathbf u$. In this approach, a vortex core corresponds to a
pressure minimum not induced by viscous effects nor unsteady straining. It is defined
as a connected region with two negative eigenvalues of $\bf T$. A vortex is therefore
detected at a given location in the fluid domain if the median eigenvalue, denoted
by $\lambda_2$, is locally negative (\cite{jeong1995_jfm}).
%
\section{Steady flow regimes at low Reynolds number \label{sec:steady}}
\begin{figure}
\psfrag{sn1}{SN$_1$}
\psfrag{sn2}{SN$_2$}
\psfrag{sn3}{SN$_3$}
\psfrag{sn4}{SN$_4$}
\psfrag{sn5}{SN$_5$}
\psfrag{sn6}{SN$_6$}
\psfrag{f1}{F$_1$}
\psfrag{f2}{F$_2$}
\psfrag{f3}{F$_3$}
\psfrag{0}{0}
\psfrag{0.563}{0.563}
\psfrag{0.366}{0.366}
\begin{center}
\includegraphics[width=0.5\textwidth]{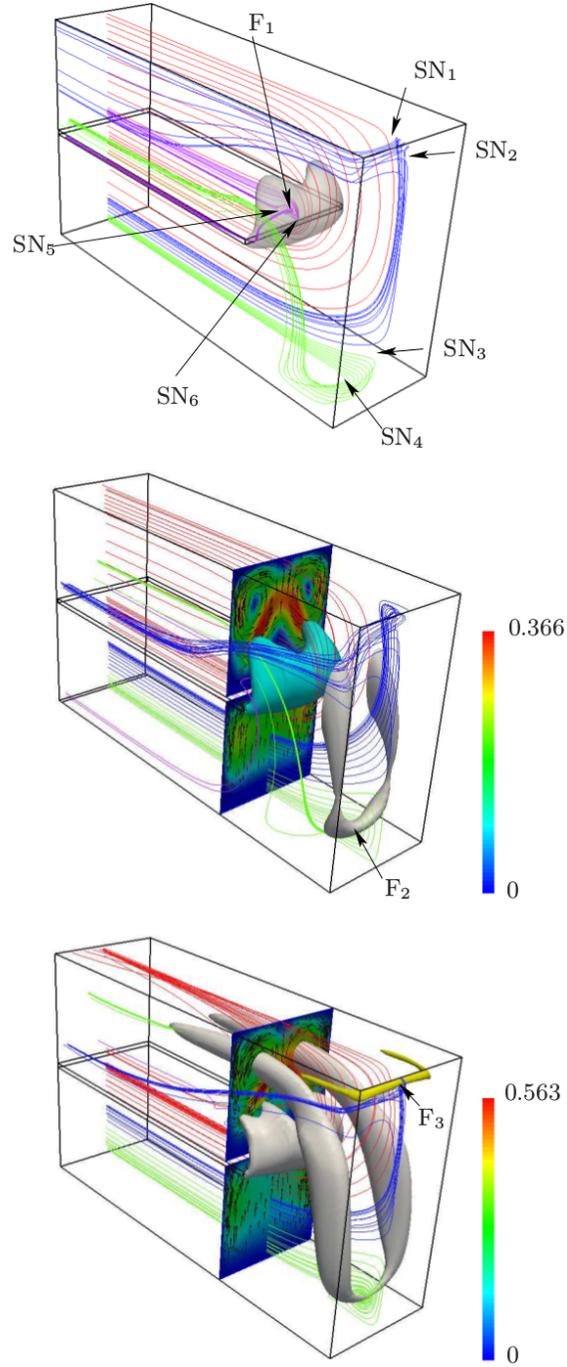}
\end{center}
\caption{Three-dimensional streamlines and iso-surfaces of $\lambda_2$ (Top) $Re=50$, $\lambda_2=10^{-4}$, (Middle) $Re=100$, $\lambda_2=4\times10^{-4}$ (cyan), $\lambda_2=5\times 10^{-5}$ (grey), (Bottom) $Re=300$, $\lambda_2=6\times10^{-4}$ (grey), $\lambda_2=6\times 10^{-4}$ (yellow). In the $(y,z)$ plane, contours and arrows respectively represent $(u_y^2+u_z^2)^{1/2}$ and $u_y\mathbf e_y + u_z\mathbf e_z$.
}
\label{fig:3dstreamlines}
\end{figure}

At very low values of $Re$ (typically 5), the flow is in a creeping state, and almost symmetric with respect to the CP, but also with respect to the $(x,0,z)$ plane. As $Re$ is increased, symmetry is lost as the flow in the turning part 
is progressively displaced towards the top outlet wall (TOP). No other change 
affects its topology until it separates from the bottom outlet wall (BOP). 
At $Re$=50, this separation is already present. The flow distortion becomes 
significant but the flow remains symmetric with respect to the CP (see 
figure \ref{fig:3dstreamlines}-(a)). 
%
%
%
%
\begin{figure}
\psfrag{a}{(a)}
\psfrag{b}{(b)}
\psfrag{c}{(c)}
\psfrag{d}{(d)}
\psfrag{sn1}{SN$_1$}
\psfrag{sn2}{SN$_2$}
\psfrag{sn3}{SN$_3$}
\psfrag{sn4}{SN$_4$}
\psfrag{sn5}{SN$_5$}
\psfrag{sn6}{SN$_6$}
\psfrag{f1}{F$_1$}
\psfrag{f2}{F$_2$}
\begin{center}
\includegraphics[width=1\textwidth]{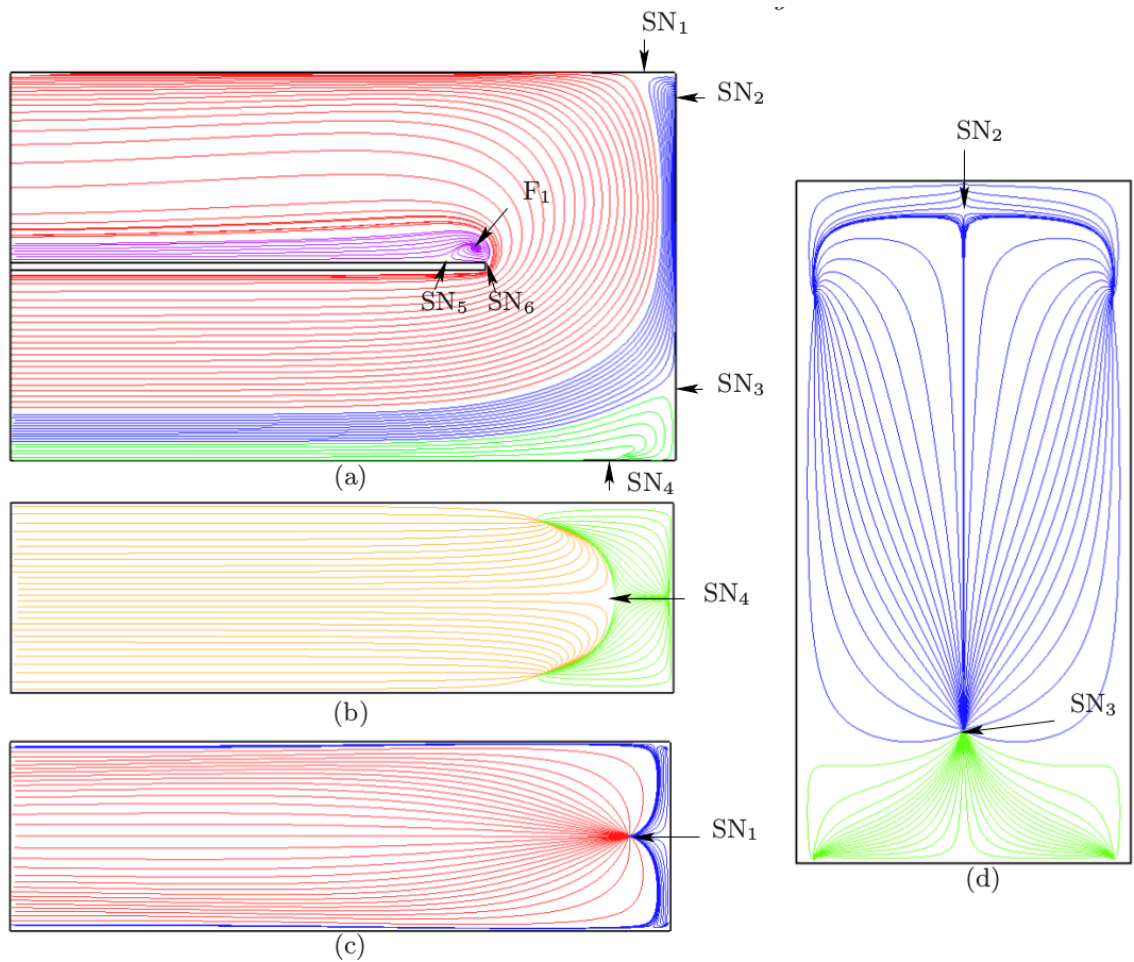}
\end{center}
\caption{Two-dimensional representation of flow patterns at $Re=50$: (a) streamlines of $(u_x,u_y)$ in the CP, and stresslines in the (b) bottom inlet plane (BIP), (c) top outlet plane (TOP) and (d) end wall (BP). 
}
\label{fig:re50_2dstreamlines}
\end{figure}
%
\subsection{Streamlines originating in the inlet centreplane ($Re=50$)
\label{sec:steady50_inlet}}
Streamlines originating in the inlet centreplane separate into three different 
streams seeded increasingly closer to the bottom inlet wall and respectively 
highlighted in red, blue, and green (see figures \ref{fig:re50_2dstreamlines} and 
\ref{fig:3dstreamlines}-(a)). 
The separation between these streams is better seen on figure. 
\ref{fig:re50_2dstreamlines}-(a)), which represents the 2D streamlines exactly in 
the CP.
All streamlines originating in the inlet follow a similar path up to the turning part (which starts at the $x=0$ plane). Red and blue lines remain in the centreplane as they follow the bend, and separate at $SN_1$ on the intersection between the CP and the TOP ($SN_1$ is a half-saddle in the CP and a node in TOP, as seen on figure \ref{fig:re50_2dstreamlines}-(c)). While red lines remain in the CP up to the outlet, blues lines head towards the end wall (or, back plane, BP) and turn to the side wall near $SN_2$, a half-node acting as a sink in the CP, and a saddle in the back Plane (figure \ref{fig:re50_2dstreamlines}-(d)). Blue lines then 
remain nearly parallel to the side walls and close to them up to the outlet.
A similar separation takes places between blue and green lines at $SN_3$, a half-saddle in the CP and a node in the BP. Green streamlines first turn downwards along the BP and then along the Bottom Inlet Plane (BIP). They then stir away 
from the CP near $SN_4$, a saddle in the BIP and half-node in the CP. The closer
 green lines originate to the BIP, the further they turn away from the BP. In 
the lower part of the inlet, they head directly towards the vicinity of $SN_4$, 
where they stir away from the CP.
After turning away from the CP near $SN_4$, green lines impact the bottom of 
the side plane, along which they turn back up to rejoin the outlet, below the 
blue streamlines (best seen on figure \ref{fig:3dstreamlines}-(a)). 

\subsection{Streamlines exiting in the outlet centreplane, forming the recirculating bubble} 
Out of all streamlines originating in the inlet CP, only red streamlines remain 
within it up to the outlet. Conversely, one group of streamlines, represented in 
purple in figures \ref{fig:re50_2dstreamlines}-(a) and 
\ref{fig:3dstreamlines}-(a), leaves the fluid domain within, or very close 
to the CP, but originates outside it. These originate from the top corners of 
the inlet section and stir towards the CP as soon a they enter the outlet part 
of the bend. They reach the CP at focus $F_1$, which 
acts as a source for an anti-clockwise spiral hitting the bottom outlet plane 
(BOP) at $SN_5$. $SN_5$, a half-saddle in the CP and a node in the BOP, 
separates purple streamlines heading directly downstream to the outlet, from 
those heading upstream to the leading edge of the BOP. This substream of 
purple lines is separated from the mainstream (in red) by half-saddle $SN_6$ (a half saddle in CP and half-Node in the BOP) and returns to the outlet just over the spiral around $F_1$.
This structure forms the first recirculation bubble attached to the leading 
edge of the outlet part of the bend. As for other classical flows in complex 
geometry confined in all non streamwise directions (flow around obstacles 
(\cite{dp10_jfm}), behind a step (\cite{armaly1983_jfm})), no closed streamline 
exists and recirculations exchange fluid between the centre and the side of the 
duct. By contrast, in the absence of side walls, \cite{zp2013_pf} showed that 
the outlet recirculation was exclusively formed of closed streamlines. The 
spiral structure of the bubble in the presence of walls can be explained by 
Ekman pumping: since spanwise vortices rest against side walls and rotate along 
the direction normal to them $\mathbf e_z$, a flow component along $\mathbf e_z$ 
pointing away from the wall is then induced out the Bodew\"adt boundary layer 
that develops along the wall (\cite{pedlosky87}). The direction 
of the punping can also be reversed if the swirl is sufficiently inhomogeneous 
along $\mathbf e_z$ (\cite{prcd13_epje}).

\subsection{Relation between numbers of Nodes and Saddles}
From Eq (2.16) in \cite{hunt1978_jfm}, the number of saddles $\Sigma^{(\mathcal S)}_S$ and nodes $\Sigma^{(\mathcal S)}_N$  formed by streamlines in a $n-$connected two-dimensional surface $\mathcal S$ must satisfy:
\begin{equation}
\Sigma^{(\mathcal S)}_N-\Sigma^{(\mathcal S)}_S=1-n.
\label{eq:hunt}
\end{equation}
In total, counting foci as nodes, streamlines in the CP form $\Sigma^{(CP)}_N=1+2\times(1/2)$ nodes and $\Sigma^{(CP)}_S=4\times(1/2)$ saddles. The fluid domain 
within the centreplane being simply connected, $n=1$, so $\Sigma^{(CP)}_N-\Sigma^{(CP)}_S=0$ thus satisfies (\ref{eq:hunt}). Two points should be underlined concerning the critical point analysis:
\begin{enumerate}
\item Moffatt vortices (\cite{moffatt1964_jfm,hancock1981_jfm}) generate an infinite number of focii and saddles in corner regions. These can be safely ignored on the basis that each of these vortices adds one focus and one saddle and hence satisfies (\ref{eq:hunt}) locally. 
\item The recirculating bubble near the outlet forms one focus and two half-saddles in the centreplane, and thus does satisfy (\ref{eq:hunt}).  Hence a flow featuring the complex flow structure within the turning part but without recirculation on the outlet part would be topologically consistent. In this sense, these two parts of the flow topology are independent.
\end{enumerate}
%
\section{Secondary flows at higher Reynolds numbers} 
As $Re$ is increased in the steady flow regime ($Re<800$), two important 
changes in the topological structure of the flow take place, first in the 
bottom corner of the turning part, then in the top corner. At the same time, 
the position of some of the critical points identified in the previous section 
vary and so does their relative importance for the overall dynamics. 
\subsection{Dean vortices originating in the bottom corner of the turning part}
%
\begin{figure}
\begin{center}
\psfrag{a}{(a)}
\psfrag{b}{(b)}
\psfrag{c}{(c)}
\psfrag{d}{(d)}
\psfrag{sn1}{SN$_1$}
\psfrag{sn2}{SN$_2$}
\psfrag{sn3}{SN$_3$}
\psfrag{sn4}{SN$_4$}
\psfrag{sn5}{SN$_5$}
\psfrag{sn6}{SN$_6$}
\psfrag{f1}{F$_1$}
\psfrag{f2}{F$_2$}
\includegraphics[width=\textwidth]{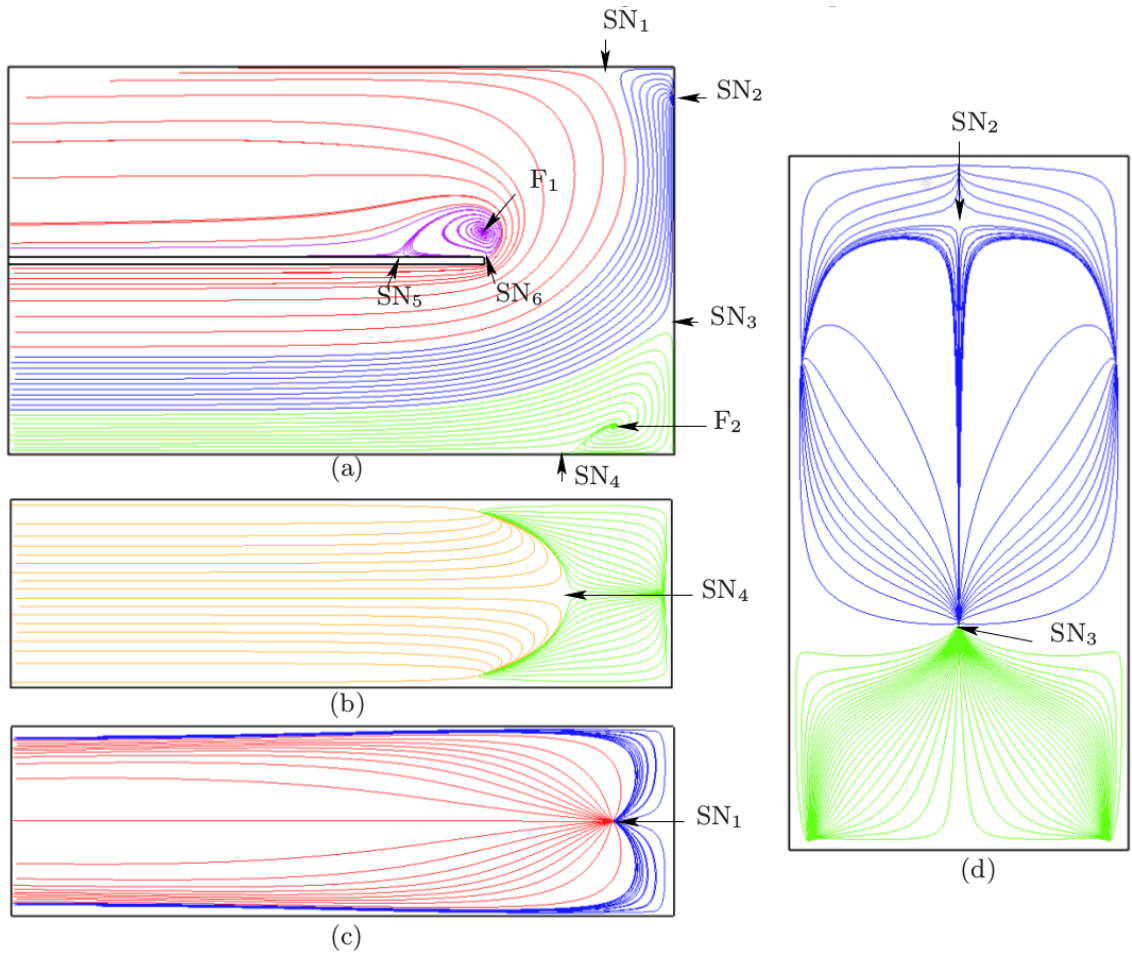}
\caption{Two-dimensional representation of flow patterns at $Re=100$: (a) streamlines of $(u_x,u_y)$ in the CP, and stresslines in the (b) bottom inlet plane (BIP), (c) top outlet plane (TOP) and (d) end wall (BP).
}
\label{fig:re100_2dstreamlines}
\end{center}
\end{figure}
\begin{figure}
\psfrag{os}{Outlet side}
\begin{center}
\psfrag{hhh1}{$h_1$}
\psfrag{hhh2}{$h_2$}
\psfrag{hhh3}{$h_3$}
\psfrag{hhh4}{$h_4$}
\psfrag{hhh5}{$h_5$}
\psfrag{h}{$h/a$}
\psfrag{Re}{$Re$}
\includegraphics[width=0.8\textwidth]{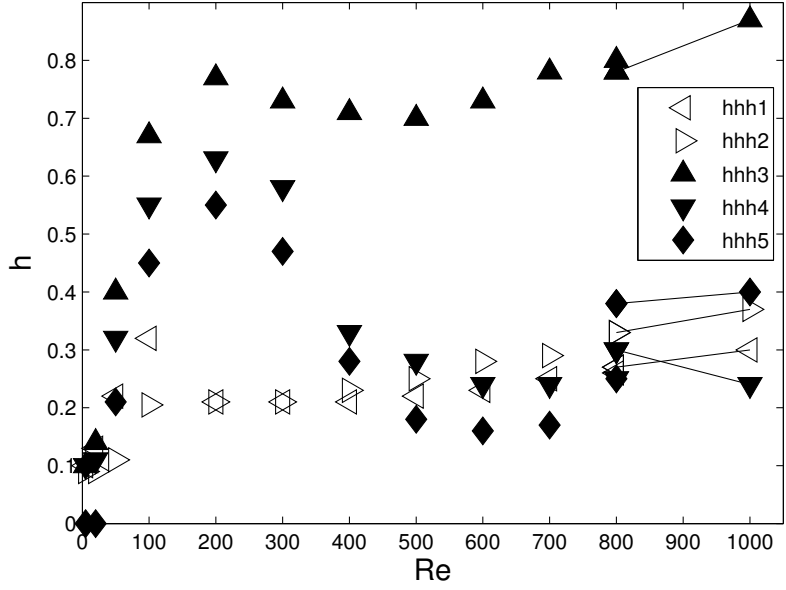}
\end{center}
\caption{Variation of the position of the main critical points in the CP. $h_1,h_2$: distance from $SN_1$ and $SN_2$ to top corner, $h_3,h_4$: distance from $SN_3$ and $SN_4$ to bottom corner. $h_5$: distance from $SN_5$ to BOP leading edge. Points joined by a line correspond to unsteady regimes (For Re=800, the steady point correspond to the long, apparently steady phase and the unsteady point corresponds to the periodic phase). 
}
\label{fig:hwithre}
\end{figure}

At $Re=100$ (figures \ref{fig:3dstreamlines}-(b) and 
\ref{fig:re100_2dstreamlines}), green lines do not converge towards $SN_3$ 
anymore but whirl around a new focus point $F_2$ located within the bulk of 
the flow instead. In the process, $SN_4$ becomes a half-saddle in the CP but remains a saddle in the BIP. Instead of turning towards the side walls in the vicinity of $SN_4$ as they did at $Re=50$, green streamlines now form two symmetric vortex tubes connected at $F_2$. 
As $Re$ is increased beyond $Re=100$, this increasingly strong counter-rotating pair of 
streamwise vortices fed by $F_2$ extends from the turning 
part well into the outlet, where its occupies an increasingly large space on 
either sides of the the CP. The pair is identified on figure \ref{fig:3dstreamlines}-(a),(b) 
using iso surfaces of $\lambda_2$ set to $5\times10^{-5}$ at $Re=100$ and 
to $6\times10^{-4}$ at $Re=300$ (\cite{jeong1995_jfm}).
These structures are the same type of vortices as those first identified near curved boundaries by 
\cite{dean1927_pm,dean1928_pm}, and in a number of other configurations 
involving ducts and pipes with various curvatures such as 90$^o$ bends 
(\cite{humphrey1977_jfm}) and others (\cite{berger1983_arfm}). These so-called 
\emph{Dean vortices} form 
as a result of the strong curvature of the streamlines in the turning part. 
The centripetal pressure drop they induce is stronger near the CP than near the side walls, 
where the flow is weaker. The pressure imbalance induces a converging flow from the side wall region towards the centre which, in recirculating up, creates a counter 
rotating vortex pair rotating along a streamwise axis. 
An interesting feature of the Dean flow is that the streamlines forming it exist at very low Reynolds 
numbers suggesting that it does not result from an instability 
but grows progressively in intensity as $Re$ increases. However, in all 
simulations at $Re=50$ and below, focus $F_2$ is absent and the Dean flow 
degenerates into two streams following the BOP (see section 
\ref{sec:steady50_inlet}).\\
From the topological point of view, $SN_4$, a half-node at $Re=50$, becomes a
half-saddle at $Re=100$ when focus $F_2$ is created: hence, at $Re=100$, ignoring Moffatt vortices, two-dimensional streamlines in the CP form $\Sigma^{(CP)}_N=2+2\times(1/2)$ nodes and $\Sigma^{(CP)}_S=1+4\times(1/2)$ saddles. This 
confirms that the topological change we identified in relation to the 
appearance of Dean vortices is compatible with topological constraint (\ref{eq:hunt}).\\
As $Re$ increases, $F_2$ moves towards the end wall and away from the bottom wall. This displacement is 
opposite to what interaction from a point vortex located at $F_2$ with the walls would imply and 
therefore results from the increased pressure gradient in the inlet.
At the same time the position of half-saddles $SN_3$ and $SN_4$, reported on figure \ref{fig:hwithre}
also evolves. At the lowest values of $Re$, both points move away from the bottom 
corner with increasing Reynolds. However, once $F_2$ is present, the swirl it induces and 
its motion respectively oppose the displacements of $SN_3$ and $SN_4$. Since $F_2$ is closer to the bottom wall, the effect is more pronounced on the position of $SN_4$ than $SN_3$. For $Re\geq200$, the 
displacements of both $SN_3$ and $SN_4$ even reverse, to aim towards the corner. 
The intensity of the DV relative to the main stream then saturates for higher 
values of $Re$ and for $Re\geq500$, the main stream displaces $SN_3$ away from 
the bottom left corner again, while $SN_4$ remains mostly at the same position. 
\subsection{Bullhorn vortices originating in the top corner}
\begin{figure}
\begin{center}
\psfrag{a}{(a)}
\psfrag{b}{(b)}
\psfrag{c}{(c)}
\psfrag{d}{(d)}
\psfrag{sn1}{SN$_1$}
\psfrag{sn2}{SN$_2$}
\psfrag{sn3}{SN$_3$}
\psfrag{sn4}{SN$_4$}
\psfrag{sn5}{SN$_5$}
\psfrag{sn6}{SN$_6$}
\psfrag{f1}{F$_1$}
\psfrag{f2}{F$_2$}
\psfrag{f3}{F$_3$}
\includegraphics[width=\textwidth]{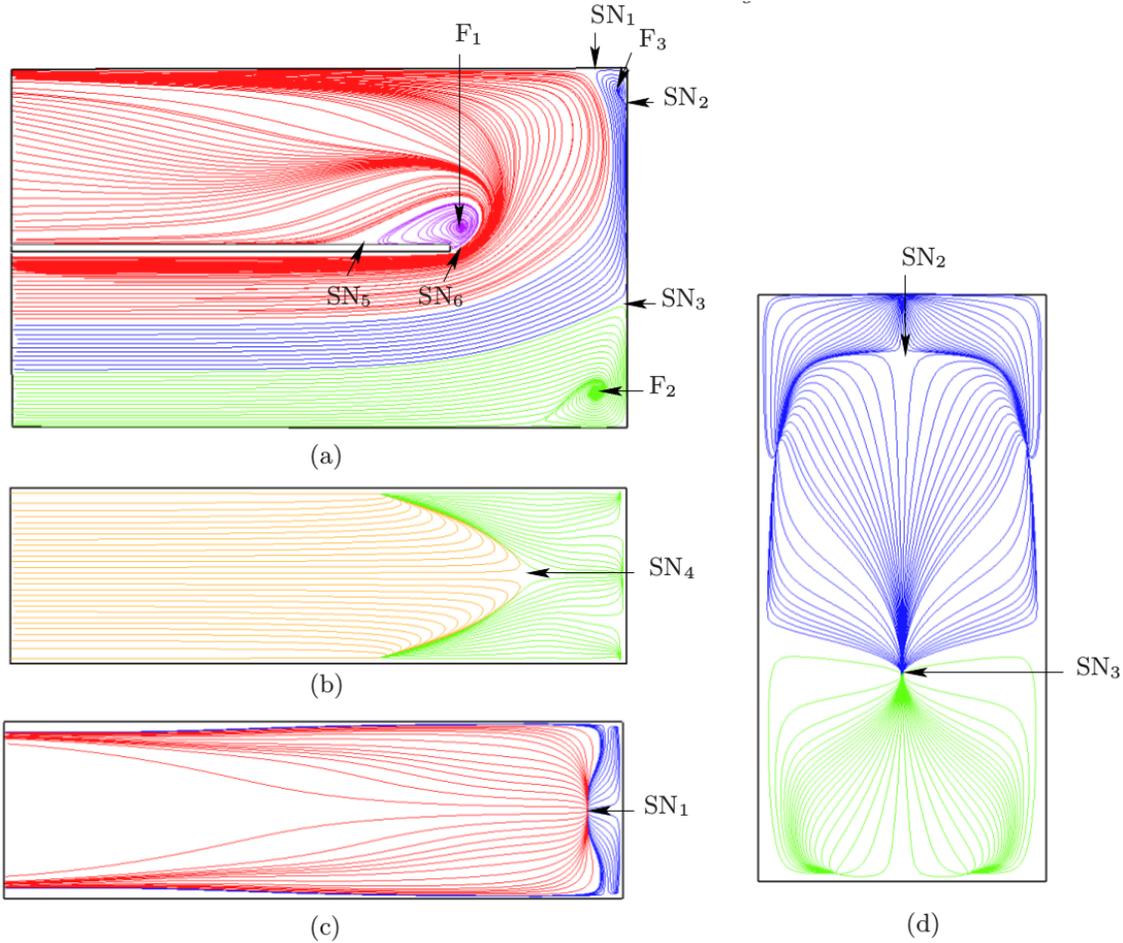}
\caption{Two-dimensional representation of flow patterns at $Re=300$: (a) streamlines of $(u_x,u_y)$ in the CP, and stresslines in the (b) bottom inlet plane (BIP), (c) top outlet plane (TOP) and (d) end wall (BP).}
\label{fig:re300_2dstreamlines}
\end{center}
\end{figure}

A second topological change identical to that affecting $SN_4$ at $Re=100$ takes
 place at $Re=300$. In turn, $SN_2$ becomes a half-saddle in the CP but remains 
a saddle in the BP, while a new focus, $F_3$ appears near the top corner of the 
turning part (See corresponding topology on figure \ref{fig:re300_2dstreamlines}). This time, blue streamlines cease to converge towards $SN_2$ and 
whirl around $F_3$ instead. They form an additional pair of counter-rotating 
vortex tubes connected at $F_3$. These are in the shape of a bull's horns, and 
extend into the outlet, along its top corners
(see figure \ref{fig:3dstreamlines}-(c)).\\
As for Dean vortices, the topological change associated to the appearance of 
bullhorn vortices (BHV) satisfies topological constraint (\ref{eq:hunt}). The swirl 
motion associated to focus $F_3$ is much weaker than for $F_2$. Nevertheless, 
it is most likely responsible for the drift of $SN_2$ away from the upper top 
corner seen at values of $Re$ for which $F_3$ is present 
(see figure \ref{fig:hwithre}). The displacement of $SN_1$ away from the same 
corner by contrast is rather driven by the main stream, which overcomes the 
influence of $F_3$.

\subsection{Impact of the secondary flows:
\label{sec:secondary_flows}}
\begin{figure}
\begin{center}
\psfrag{Re}{$Re$}
\psfrag{QQ}{$Q/(U_0 a)$}
\psfrag{Q}{$z$}
\psfrag{ux}{$u_x(x_{\tiny \rm DV},0,z)$}
\includegraphics[width=0.49\textwidth]{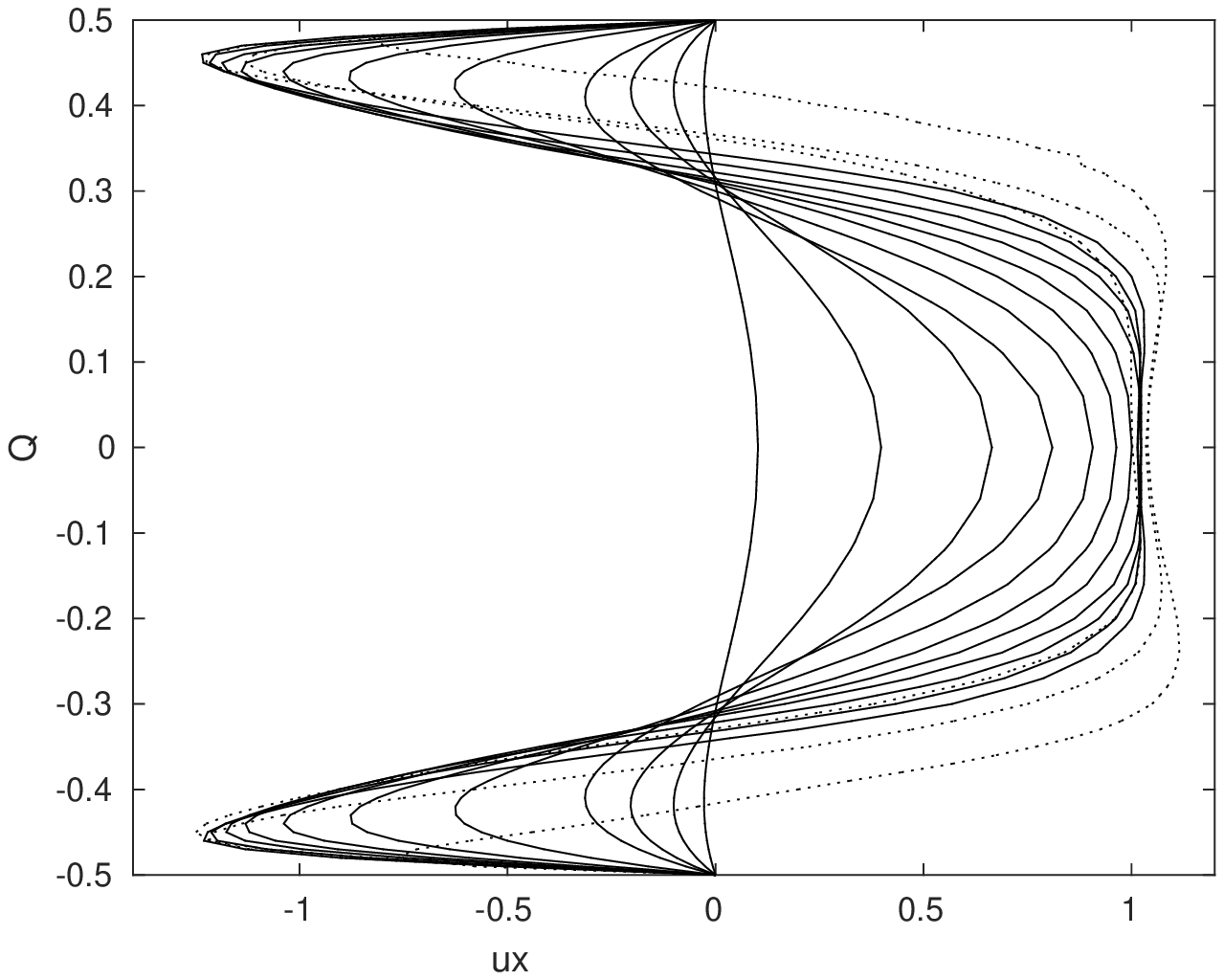}
\psfrag{z}{$z$}
\psfrag{uy}{$u_y(0.8,y_{\tiny \rm BHV},z)$}
\includegraphics[width=0.49\textwidth]{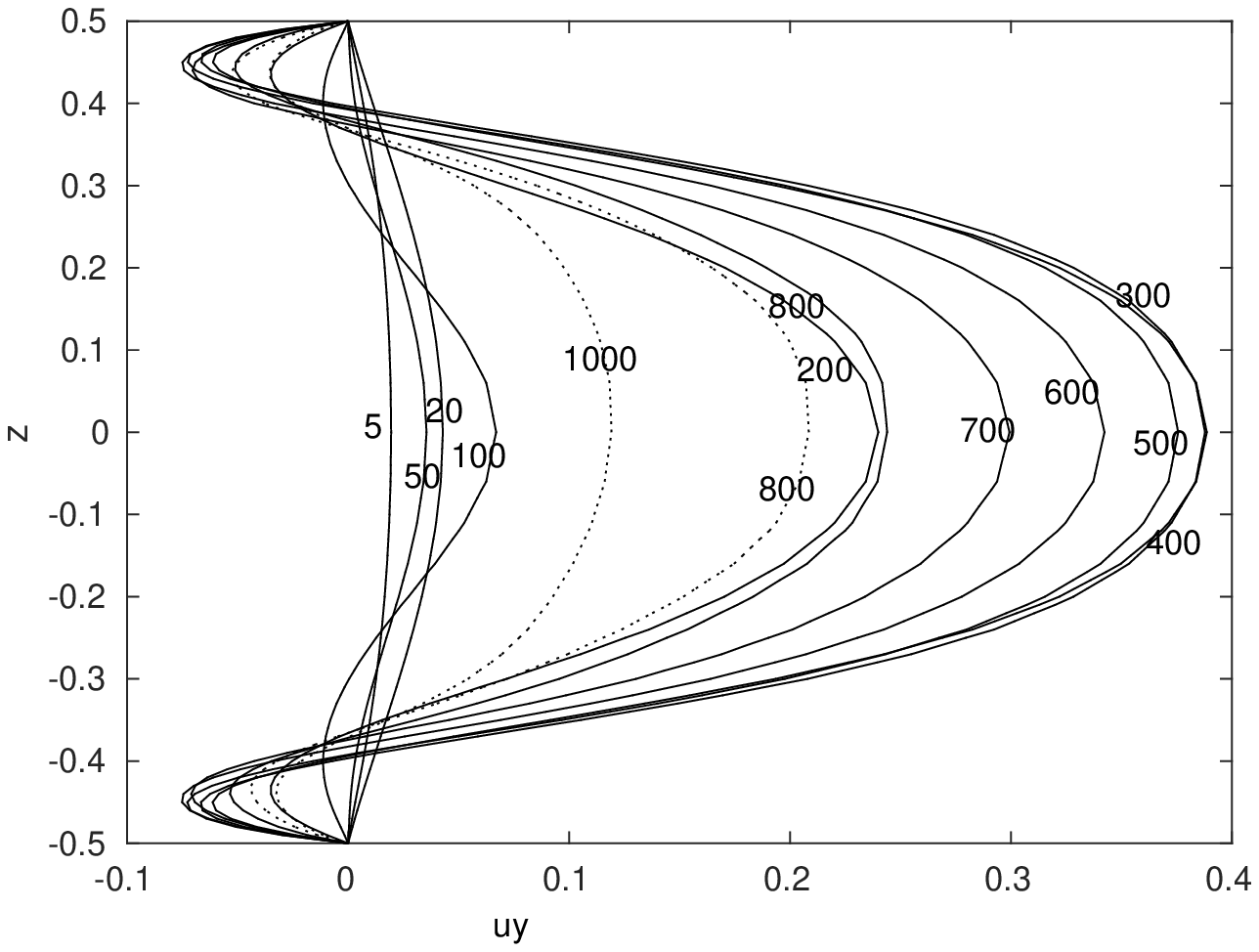}\\
\psfrag{Q}{\tiny Q}
\includegraphics[width=0.7\textwidth]{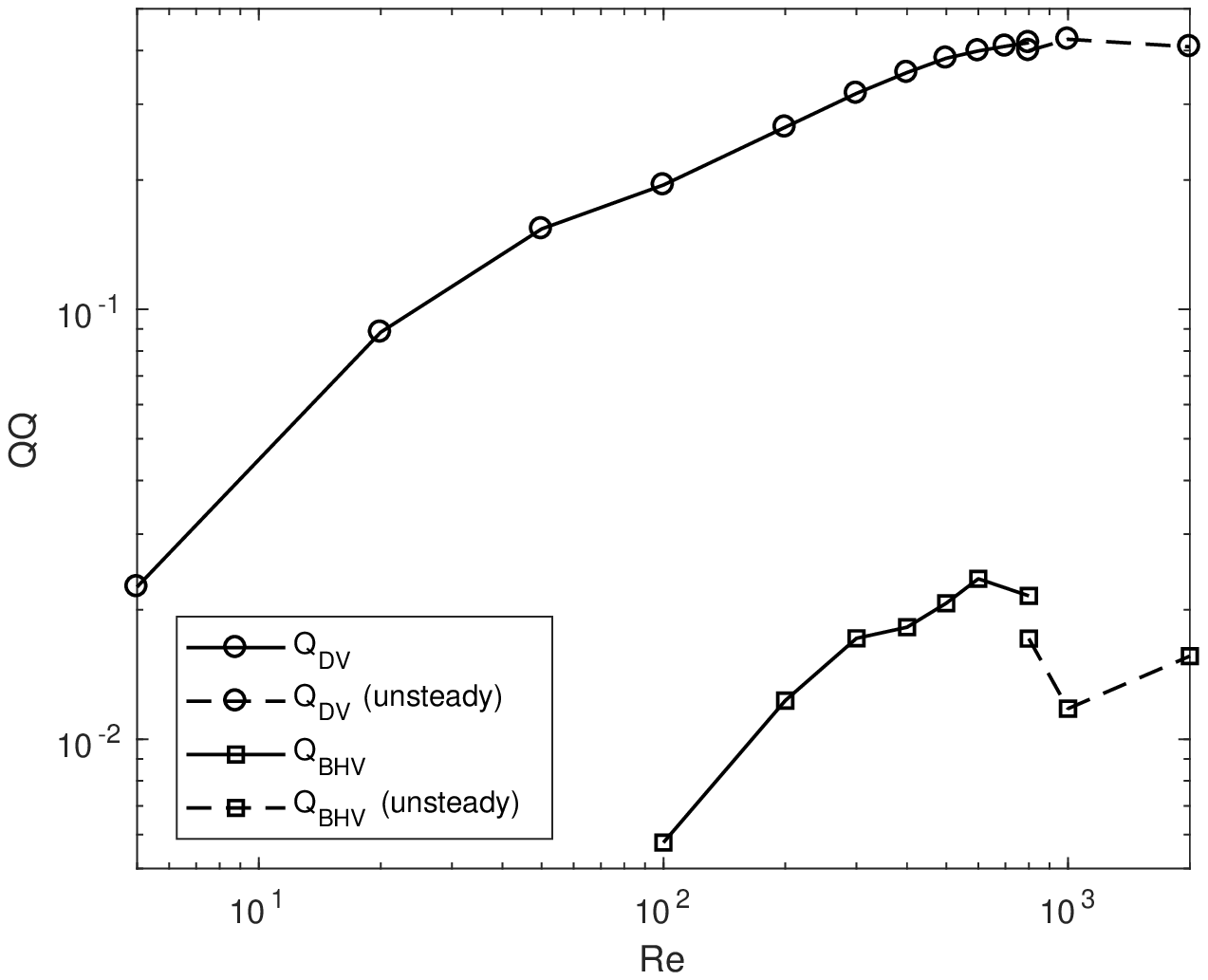}

\end{center}
\caption{Top left: velocity profile associated to the Dean Flow in the middle of the turning part 
$u_x(x_{\rm DV},y_{\rm DV}, z), z)$. 
Top right: Velocity profile associated to the Bullhorn vortices in the middle 
of the turning part $u_x(x_{\rm BHV},y_{\rm BHV}, z), z)$.
Solid line, steady flows for $Re=$5, 20, 50, 100, 200, 300, 400, 500, 600, 700 and 800. Dotted line, unsteady flows for $Re=$800, 1000 and 2000. Whilst for DV profiles the order of growing maximum velocities follows the values of $Re$ within the steady and unsteady regimes, note the non-trivial order of these maxima for BHV. This reflects a suppression of the BHV for $Re\gtrsim$. 
 Bottom; 2D flow rate associated to the integral of these profiles, normalised by the same quantity associated to the inflow: these non-dimensional quantities give a measure of the flow associated to the 
DV and BHV, relative to the inflow.
\label{fig:q_dv_bhv}
}
\end{figure}

\begin{figure}
\begin{center}
\psfrag{a}{(a)}
\psfrag{b}{(b)}
\psfrag{c}{(c)}
\psfrag{0}{0}
\psfrag{1.33}{1.33}
\psfrag{dv}{DV}
\psfrag{bhv}{BHV}
\includegraphics[width=\textwidth]{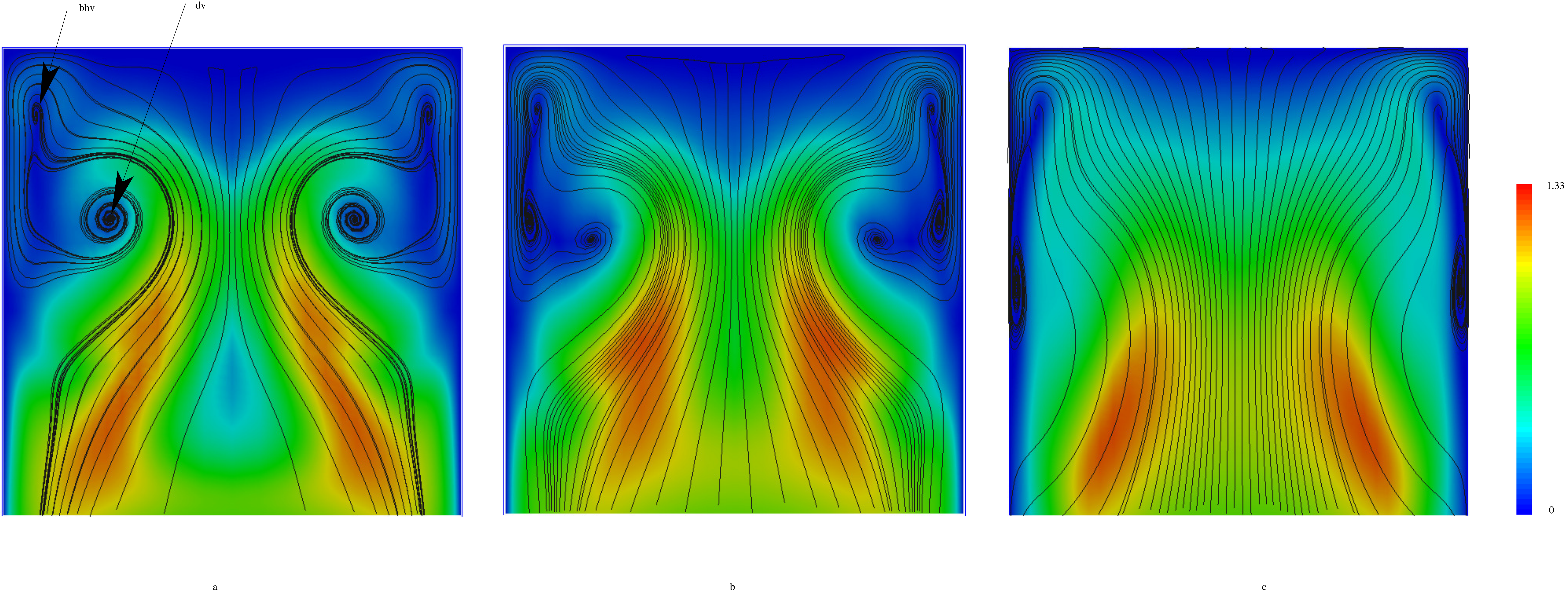}
\end{center}
\caption{Two-dimensional projection of streamlines in the outlet into planes (a) $x$=-0.08, (b) $x$=-0.07 and (c) $x$=-0.05, viewed from the end wall at $Re=600$. Colours represent $(u_y^2+u_z^2)^{1/2}$. Dean and Bullhorn vortices are respectively labelled DV and BHV on the left part of figure (a).}
\label{fig:yz_planes}
\end{figure}

The importance of the secondary flows (Dean and Bullhorn vortices) can be evaluated through their intensity relative to the main stream. The Dean vortices
 induce a strong velocity component along $\mathbf e_x$ in the middle of the 
turning part.  In their absence, only a weak contribution to this component would arise from the 
asymmetry of the mean flow with respect to the $(x,0,z)$ plane. 
The corresponding flow profiles are shown on figure \ref{fig:q_dv_bhv} (top left) and show two interesting features: the centres of the Dean vortices remain practically at the same location along $\mathbf e_z$ as $Re$ increases in the steady regime and near the onset of unsteadiness. For $Re\gtrsim300$, the profiles flatten significantly under the influence of the end wall.
The intensity of the Dean vortices is well measured by the positive part 
of the flow rate of $u_x$ through a line intercepting  the axis of the DV, arbitrarily chosen at $(x=-x_{\rm DV},y=0)$): 
\begin{equation}
Q_{\rm DV}=\frac12\left(\int_{-1/2}^{1/2} |u_x(x_{\rm DV},0,z)| dz -\int_{-1/2}^{1/2} u_x(x_{\rm DV},0,z) dz \right),
\label{eq:qdv}
\end{equation}
where $x_{DV}$ corresponds to the symmetric points where the DV axes intercept the $y=0$ plane.
%
This quantity is reported on figure \ref{fig:q_dv_bhv} (bottom). As expected, the secondary flow increases with $Re$.
When $F_2$ and the fully developed DV are present, it rapidly reaches about one third of the inlet flowrate. For $Re\geq500$, it saturates as the DV are subject to significant friction along the end wall and the side walls in the turning part. 
The topological impact of the Dean vortices in the saturated regime is best 
seen on two-dimensional streamlines of $(u_y,u_z)$ in spanwise planes, where the DV form two clear counter-rotating structures (figure \ref{fig:yz_planes}). The upper part of the outlet is clearly dominated by the DV inducing a very strong vertical flow near the CP. This flow interacts with the recirculating bubble in the lower part of the outlet, which, as a result, becomes split into two lobes located 
either side of the CP. As $Re$ is further 
increased both lobes lengthen along the streamwise direction, with a maximum length near the lateral walls rather than at the centre of the duct (see contours of $\lambda_2$ in fig. \ref{fig:3dstreamlines} ). \\
Similarly, the intensity of the Bullhorn vortices can be measured by calculating the two-dimensional 
flowrate associated to the profile of $u_y$ along a line intercepting their axis, which we chose at $(x=0.8,y=y_{\rm BHV},z)$. These profiles are shown in figure \ref{fig:q_dv_bhv} (top right). Here again, the centres of the BHV remain essentially at the same position along $\mathbf e_z$ within the steady regime but unlike the DV, the profiles exhibit little evidence of any interaction with the top wall. The two-dimensional flowrate associated with these profiles is defined as
\begin{equation}
Q_{\rm BHV}=\frac12\left(\int_{-1/2}^{1/2} |u_y(0.8,y_{\rm BHV},z)| dy -\int_{-1/2}^{1/2} u_y(0.8,y_{\rm BHV},z) dy \right),
\label{eq:qdv}
\end{equation}
where $y_{BHV}$ corresponds to the symmetric points where the BHV axes intercept the $x=0.8$ plane.
The variations of $Q_{\rm BHV}$ are plotted on figure \ref{fig:q_dv_bhv} (bottom). 
The flowrate induced by the BHV increases for $Re\geq300$. Its relative 
intensity is however about 3-4 times 
lower than that of the DV and they extend much less into the outlet than the DV. Hence, their influence 
is mostly confined to the upper corner of turning part. Figure \ref{fig:yz_planes} indeed shows that they 
remain confined there.\\

%
To conclude the analysis of the steady regimes, the overall structure of the flow in steady regimes for $Re\geq300$ raises two remarks.
\begin{enumerate}
\item The apparently complex topology of the flow up to $Re=800$ 
is in fact entirely governed by three occurrences of the same topological pattern
made of streamlines spiralling to (or out of) a focus point, and originating from (or impacting) a nearby wall at a half-saddle.
\item None of the steady solutions we found showed the presence of a 
secondary recirculating bubble on the TOP. These appear at higher Reynolds 
numbers than the first recirculation in 180$^o$ sharp bends (\cite{zp2013_pf}), 
and in confined flows behind a backward-facing step (\cite{armaly1983_jfm}) when 
the geometry is infinitely extended in the spanwise direction. This suggests 
that secondary bubbles can only develop in sufficiently wide ducts. 
When present, they significantly alter the structure of the base flow 
at the onset of unsteadiness and may interfere with the corresponding instability mechanism. 
In the case of the infinitely extended 180$^o$ bend, however, unsteadiness mostly results from 
three-dimensional instabilities localised within the first recirculation bubble
 as soon as the bend opening ratios exceeds about 0.2 (\cite{shps2016_jfm}).
\end{enumerate}
\section{Unsteady flow regimes \label{sec:unsteady}} 
\subsection{Onset of unsteadiness}
\begin{figure}
\centering
\parbox{1.5cm}{$Re=800$} \\
\begin{tabular}{cc}
\parbox{0.49\textwidth}{
\psfrag{t}{$t$}
\psfrag{C}{$C_d$}
\psfrag{d}{}
\includegraphics[width=0.49\textwidth]{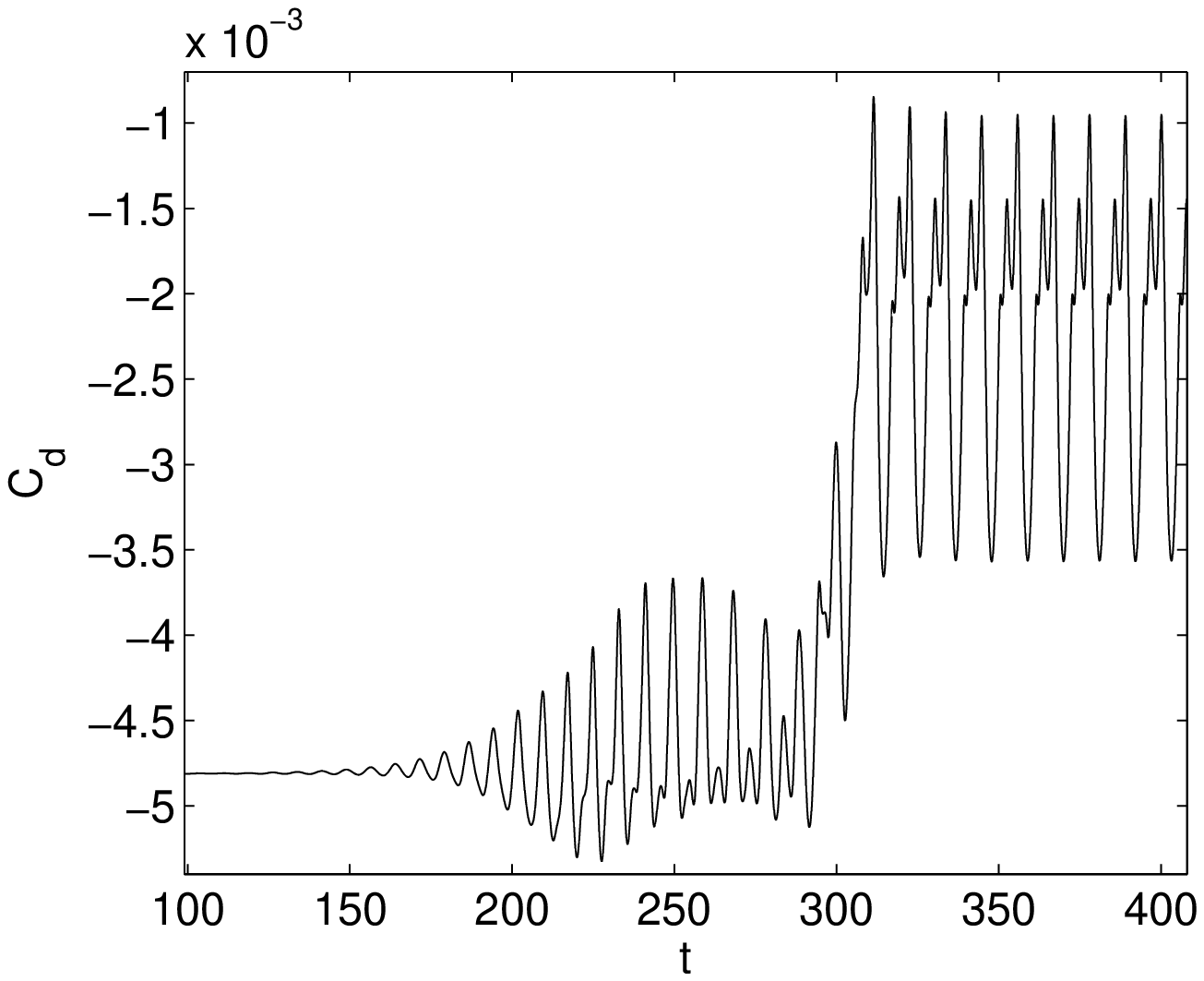}}&
\parbox{0.45\textwidth}{
\psfrag{f}{$f$}
\psfrag{FFTCd}{PSD of $C_d(t)$}
\includegraphics[width=0.49\textwidth]{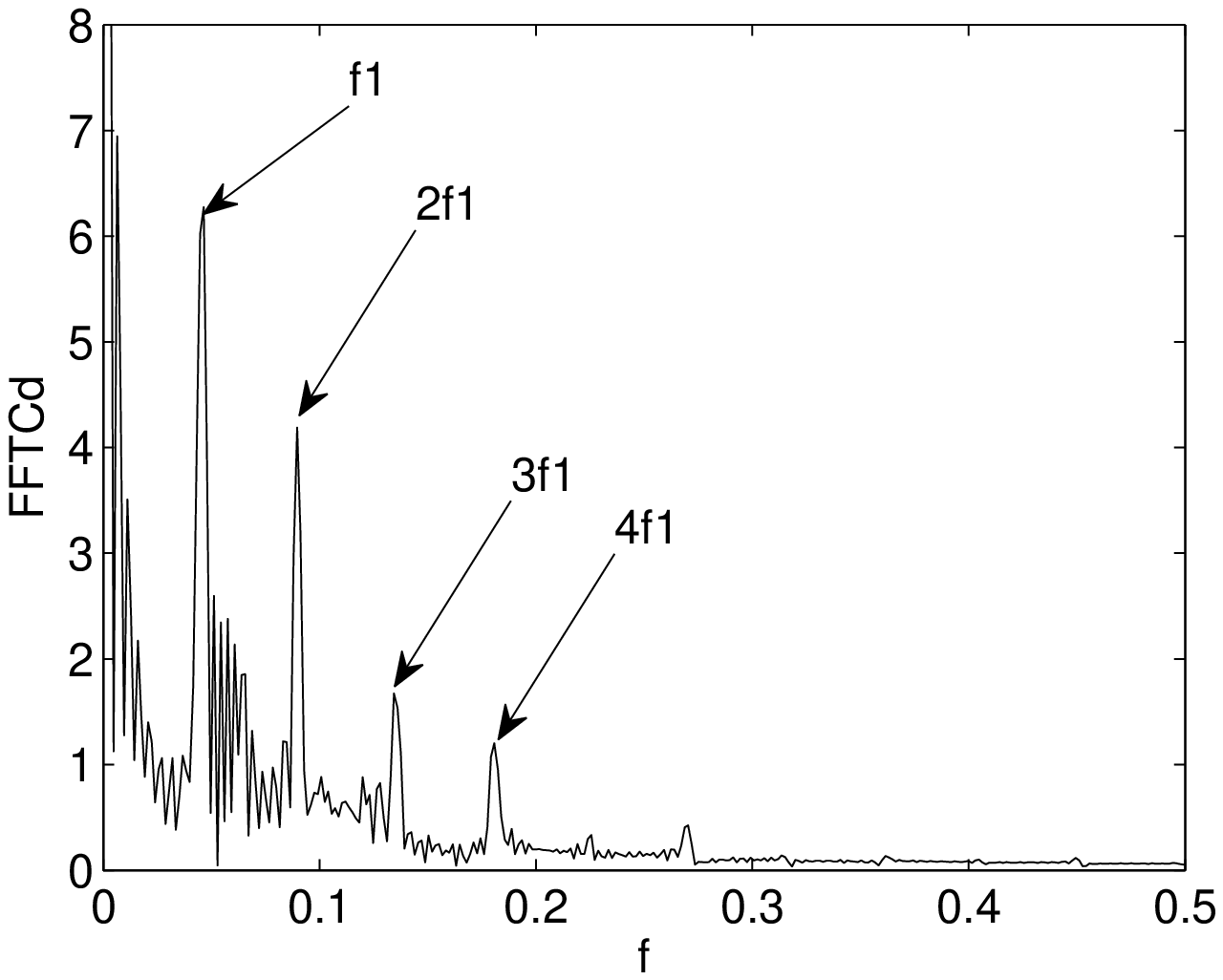}}\\
\end{tabular}
\parbox{1.5cm}{$Re=2000$} \\
\begin{tabular}{cc}
\parbox{0.45\textwidth}{
\psfrag{t}{$t$}
\psfrag{C}{$C_d$}
\psfrag{d}{}
\includegraphics[width=0.49\textwidth]{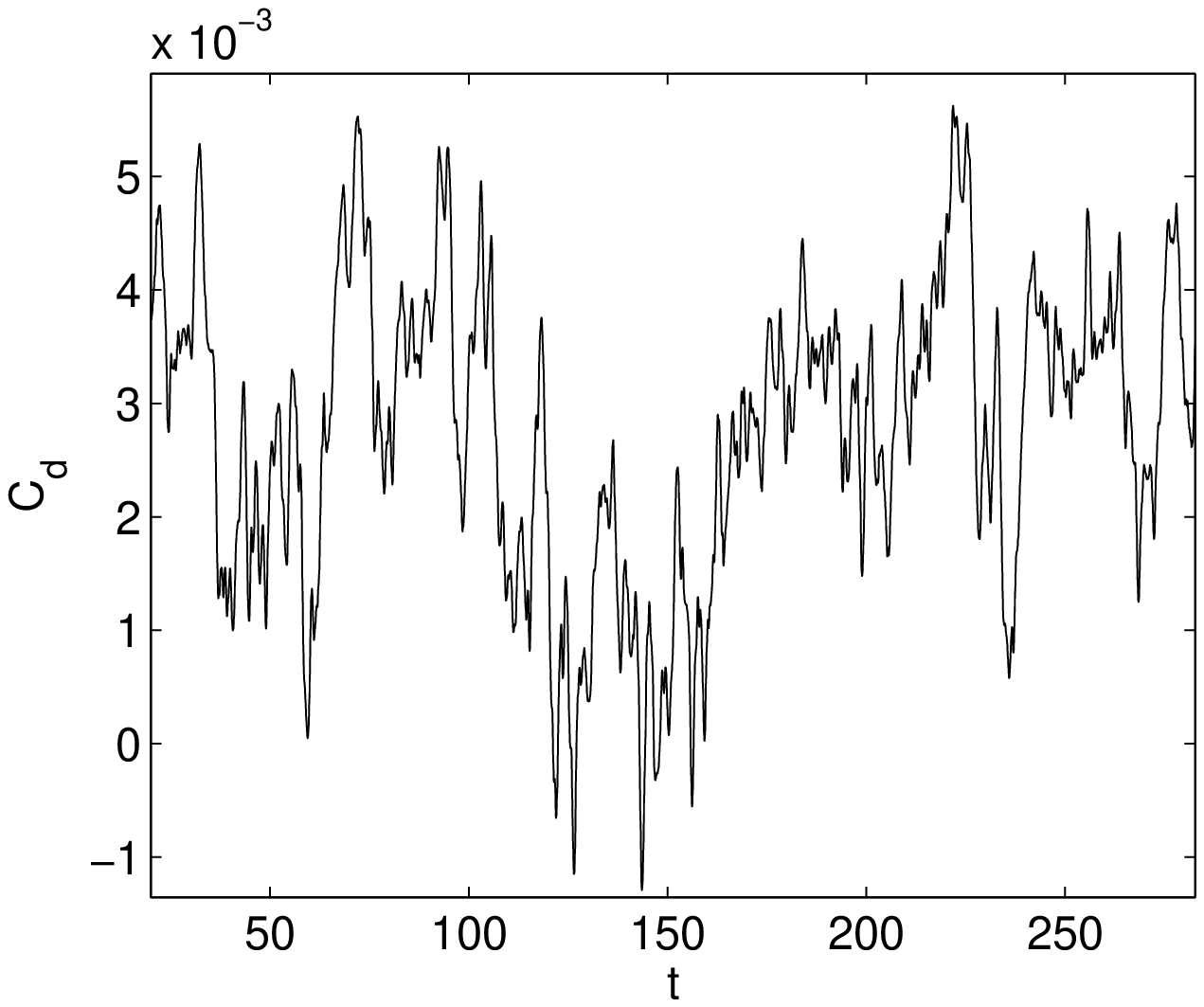}}&
\parbox{0.49\textwidth}{
\psfrag{f}{$f$}
\psfrag{FFTCd}{PSD of $C_d(t)$}
\includegraphics[width=0.49\textwidth]{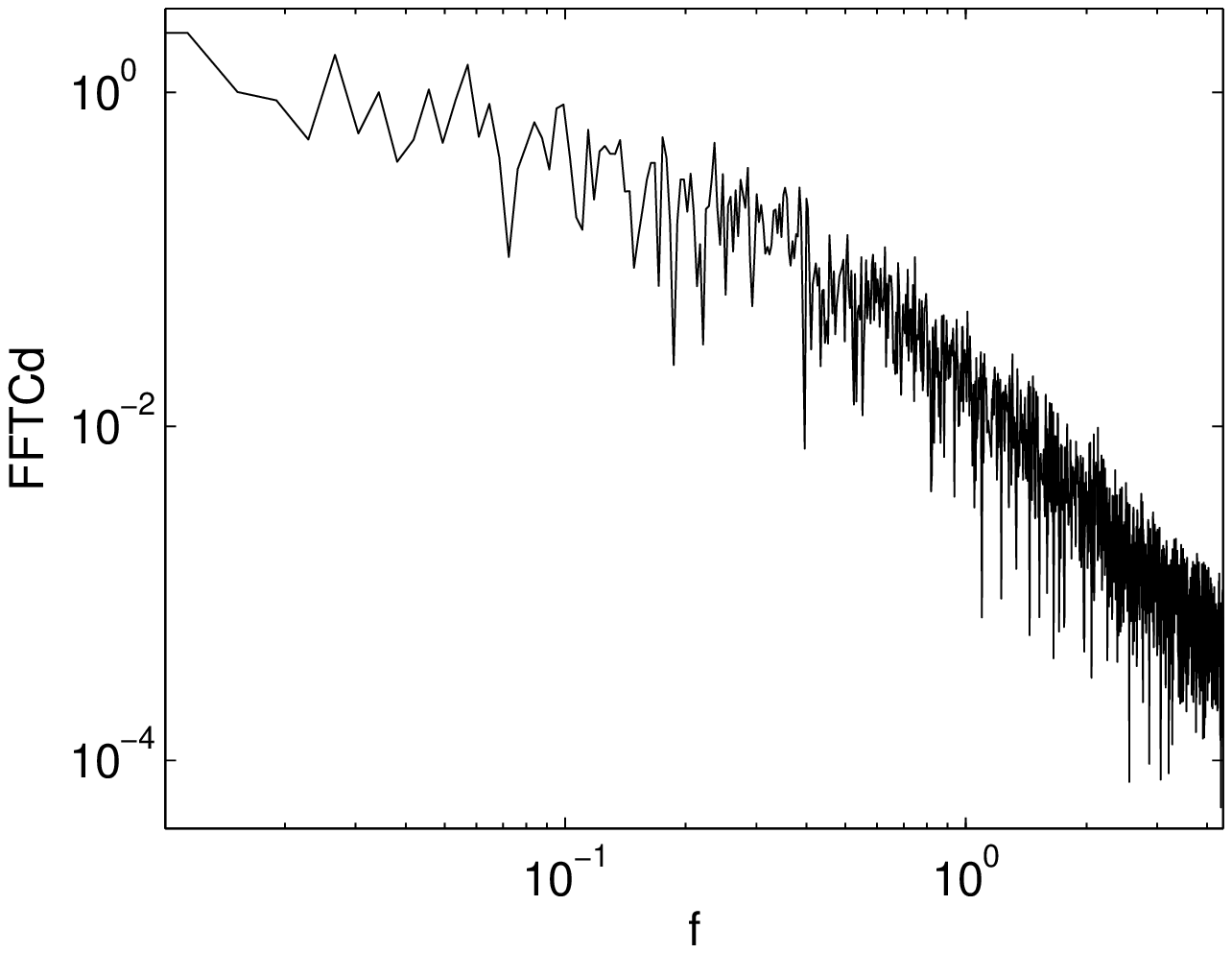}}\\
\end{tabular}
\caption{Left: time variations of the drag coefficient $C_d(t)$ on the 
bottom outlet wall. Right: Frequency Spectra obtained from time series of the drag coefficient.
\label{fig:cd}}
\end{figure}

Unsteadiness appears at Re=800 in our simulations (the highest value or Reynolds for which our simulations return a steady state is 700). It is best monitored through the drag and lift coefficients measured on the entire separating element, respectively,
\begin{eqnarray}
C_d&=&\frac1{U_0^2}\int_{BOP+TIP+ITP} \nu\partial_x u_x dS,
\label{eq:cd} \\
C_l&=&\frac1{\rho U_0^2}\int_{BOP+TIP+ITP} p dS,
\label{eq:cl}
\end{eqnarray}
where TIP and ITP are respectively the top inlet plane and the vertical plane in the inner side of the turning part.
The time variations of $C_d$ are reported on figure \ref{fig:cd}-(a), left. 
Unsteadiness first appears at $Re=800$ through a periodic oscillation of the drag of frequency 
$f_0\simeq0.135$ (calculated over the first 10 oscillations) modulated by an exponential growth. 
The oscillation fails to settle and reduces in frequency until it is interrupted by 
a brutal event taking place over $t=300-320$. The drag then suddenly settles into non-harmonic periodic 
oscillations of higher amplitude, but reduced mean. 
The established state is dominated by oscillations at a slightly lower frequency than at the 
onset $2f_1\simeq 0.089$, as well as a subharmonic frequency $f_1$,
 seen on the frequency spectrum of $C_d(t)$ (figure \ref{fig:cd}-(a), 
right).
The full spectrum also features a number of higher harmonics, of which $3f_1$ and $4f_1$ are clearly identifiable. 
 Note that our choice of non-dimensional variables makes these non-dimensional frequencies 
directly interpretable as Strouhal numbers ($St_i=\tilde f_ia/U_0$, where $\tilde f_i$ are dimensional frequencies).\\
\begin{figure}
\centering
\includegraphics[width=0.49\textwidth]{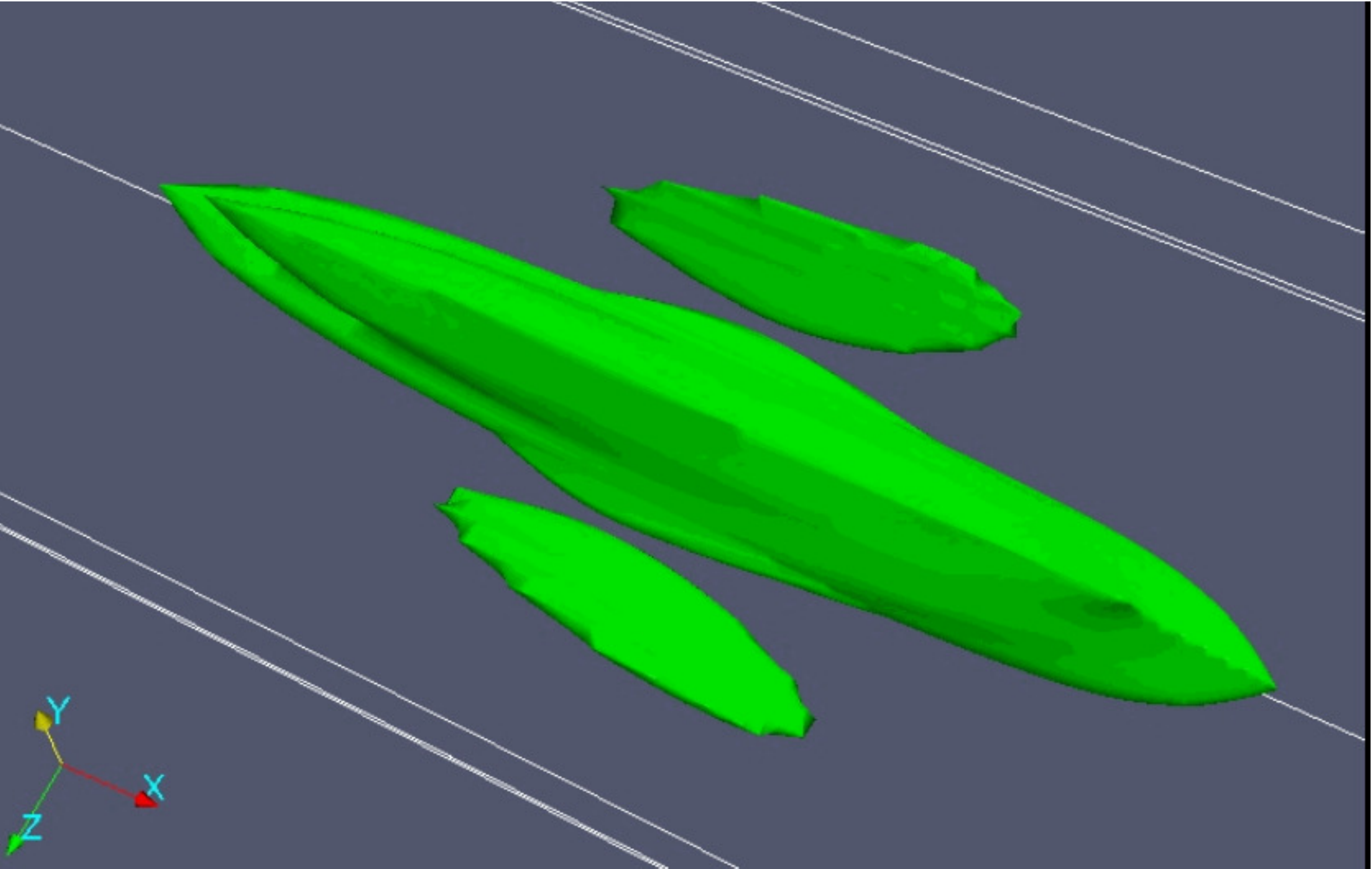}
\includegraphics[width=0.49\textwidth]{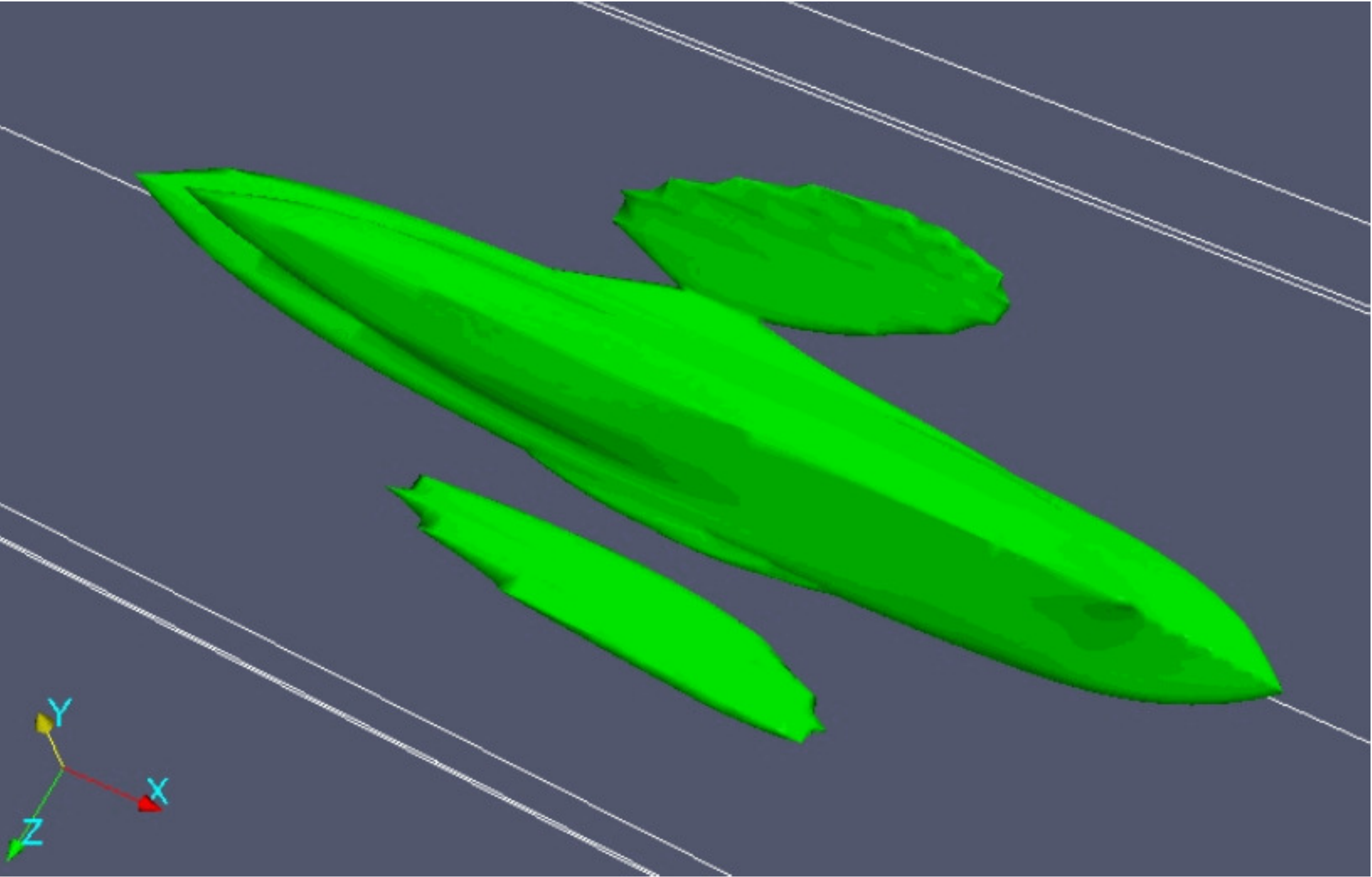}
\caption{Contours of $\lambda_2=-1.05\times10^{-3}$ showing two stages of the oscillation of the left and right lobes of the first recirculation bubble at the onset of unsteadiness. Left: symmetric position ($t=89.73$), right: asymmetric position ($t=117.33$). An animation of the oscillation is available as supplementary material in file "lambda2Re800oscillations.avi".
\label{fig:onset_lambda2}}
\end{figure}
Let us first focus on the onset phase at $t<300$. 
The corresponding mechanism can be visualised through time-dependent contours 
of $\lambda_2=-1.05\times10^{-3}$ on supplement movie "lambda2Re800oscillation.avi" and snapshots on
figure \ref{fig:onset_lambda2}: the oscillation takes its root in the alternate elongation and contraction of 
the right (\emph{i.e.} $z\geq0$) an left (\emph{i.e.} $z\leq0$) lobes of the 
recirculating bubble attached to the outlet bottom wall. 
Its origin can be understood through the action of the 
Dean flow. In steady regimes, the pair of counter-rotating Dean vortices 
present in the upper part of the outlet drives a strong jet along $\mathbf e_y$. 
Since the intensity of the pair decreases downstream of the turning part, the maximum 
intensity of this flow coincides with the location of the first recirculation bubble. 
Except for very low Reynolds numbers, the intensity of the Dean flow is sufficient to 
reshape the recirculation bubble into two almost separate lobes pushed towards the 
lateral outlet walls (section \ref{sec:secondary_flows}). Since the DV remain as strong in the unsteady 
regime (see figure \ref{fig:yz_planes}), the DV can be seen as acting to keep the lobes kinematically 
independent, thus allowing the oscillations to take place. In this sense, the interaction between 
the reciculation bubble and the DV is the root of the instability mechanism.\\
\begin{figure}
\begin{tabular}{cc}
\parbox{0.49\textwidth}{
\psfrag{t}{$t$}
\includegraphics[width=0.49\textwidth]{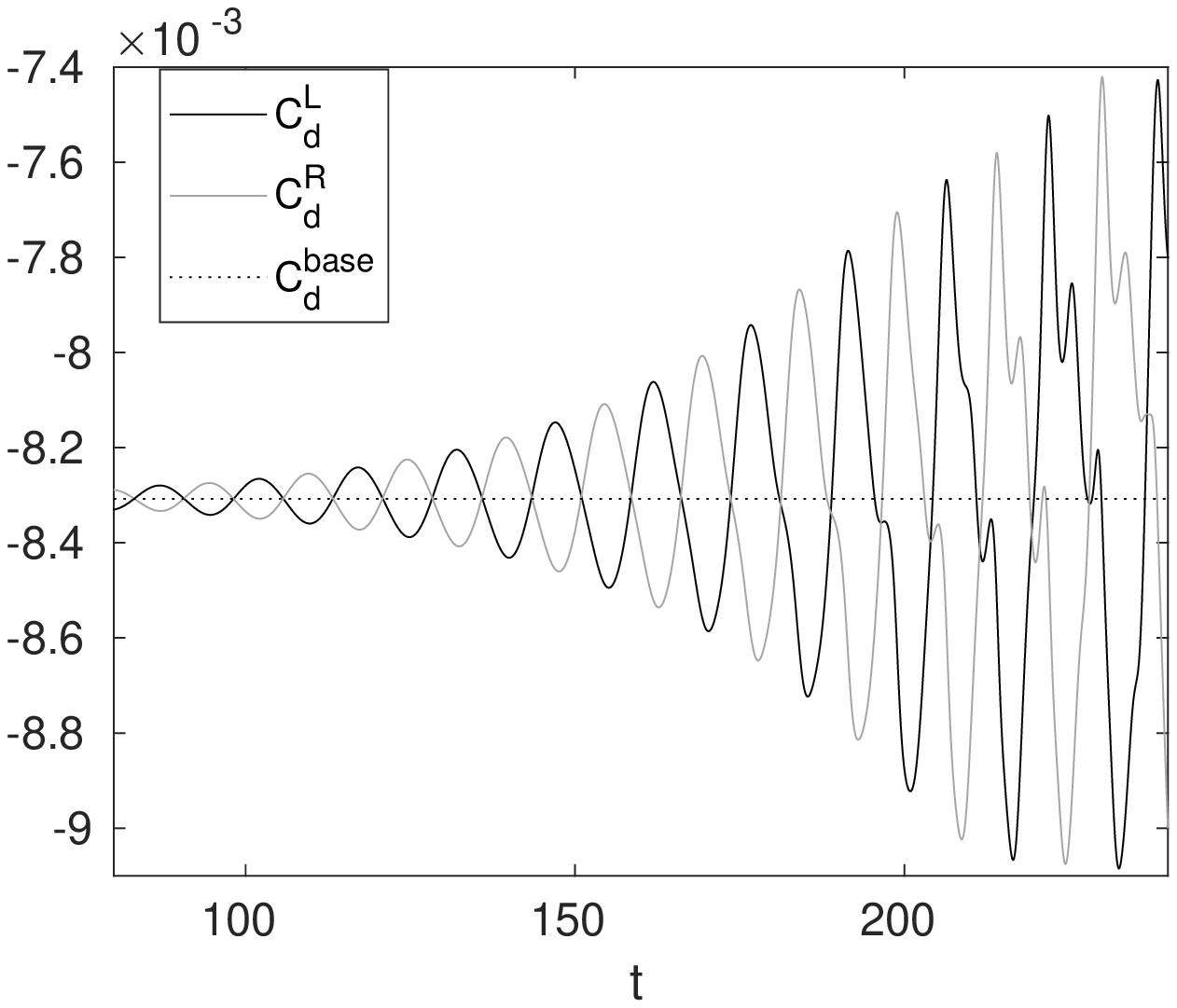}}&\parbox{0.49\textwidth}{
\psfrag{t}{$t$}
\includegraphics[width=0.49\textwidth]{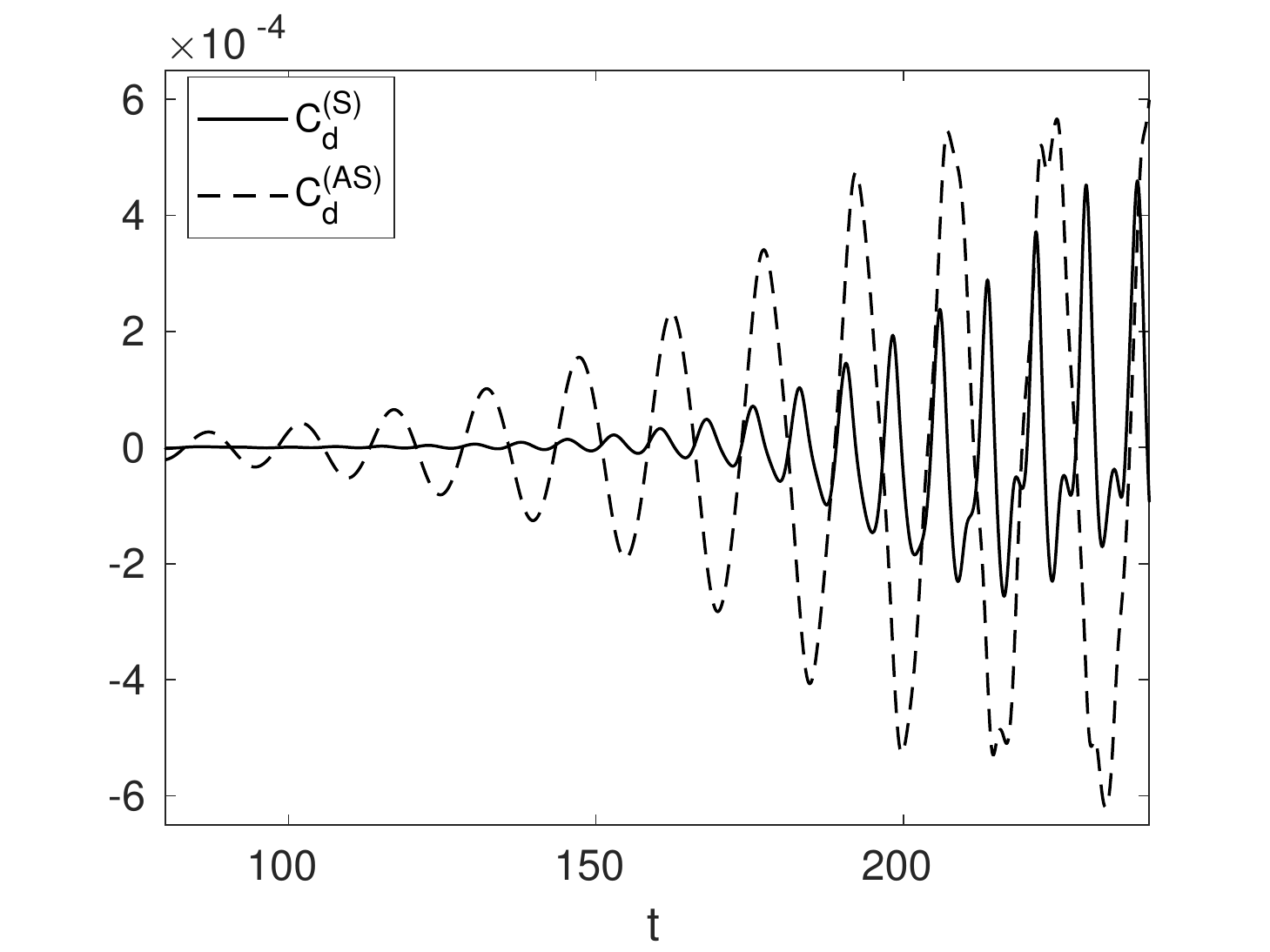}}\\
\parbox{0.49\textwidth}{
\psfrag{t}{$t$}
\includegraphics[width=0.49\textwidth]{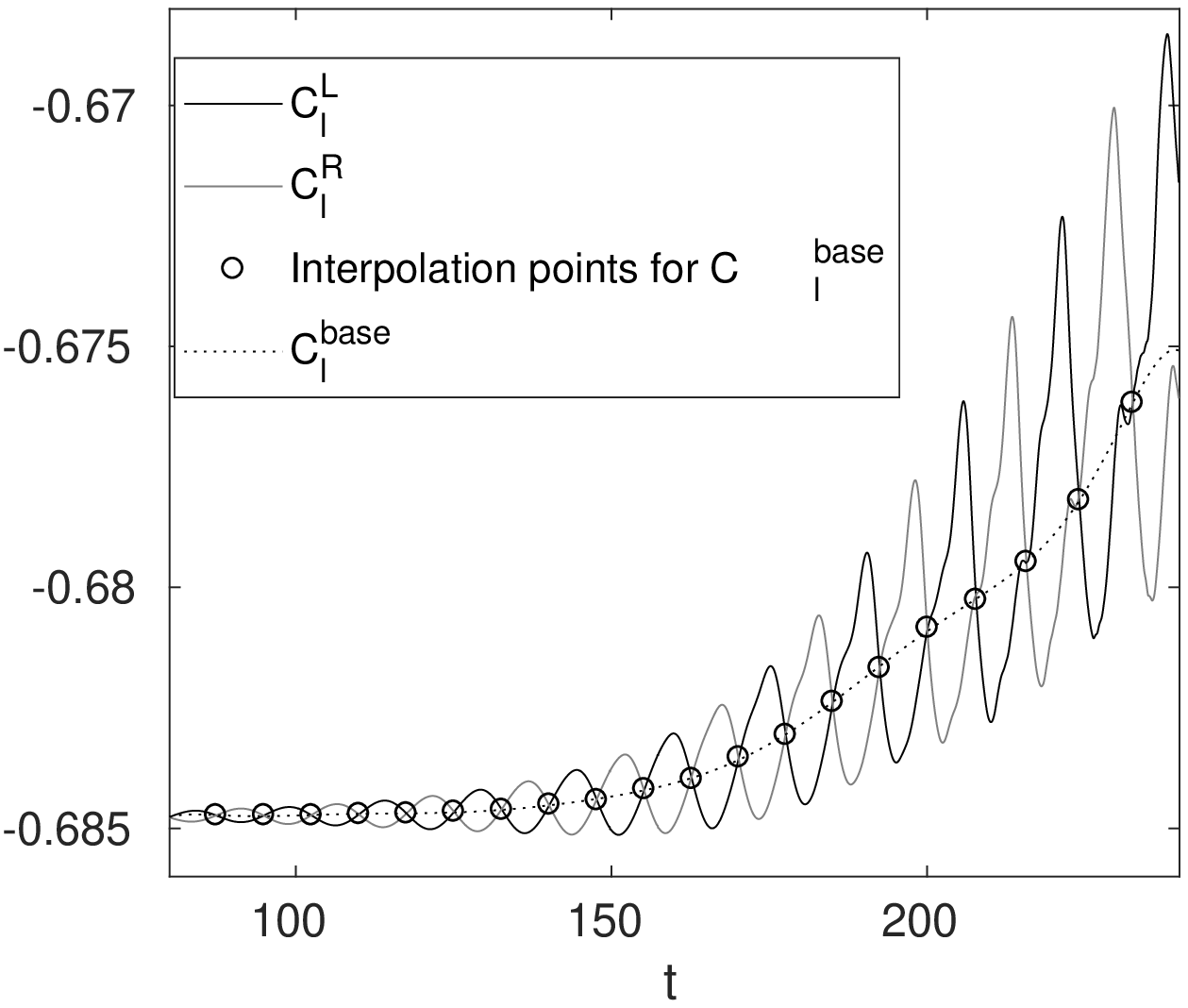}}&\parbox{0.49\textwidth}{
\psfrag{t}{$t$}
\includegraphics[width=0.49\textwidth]{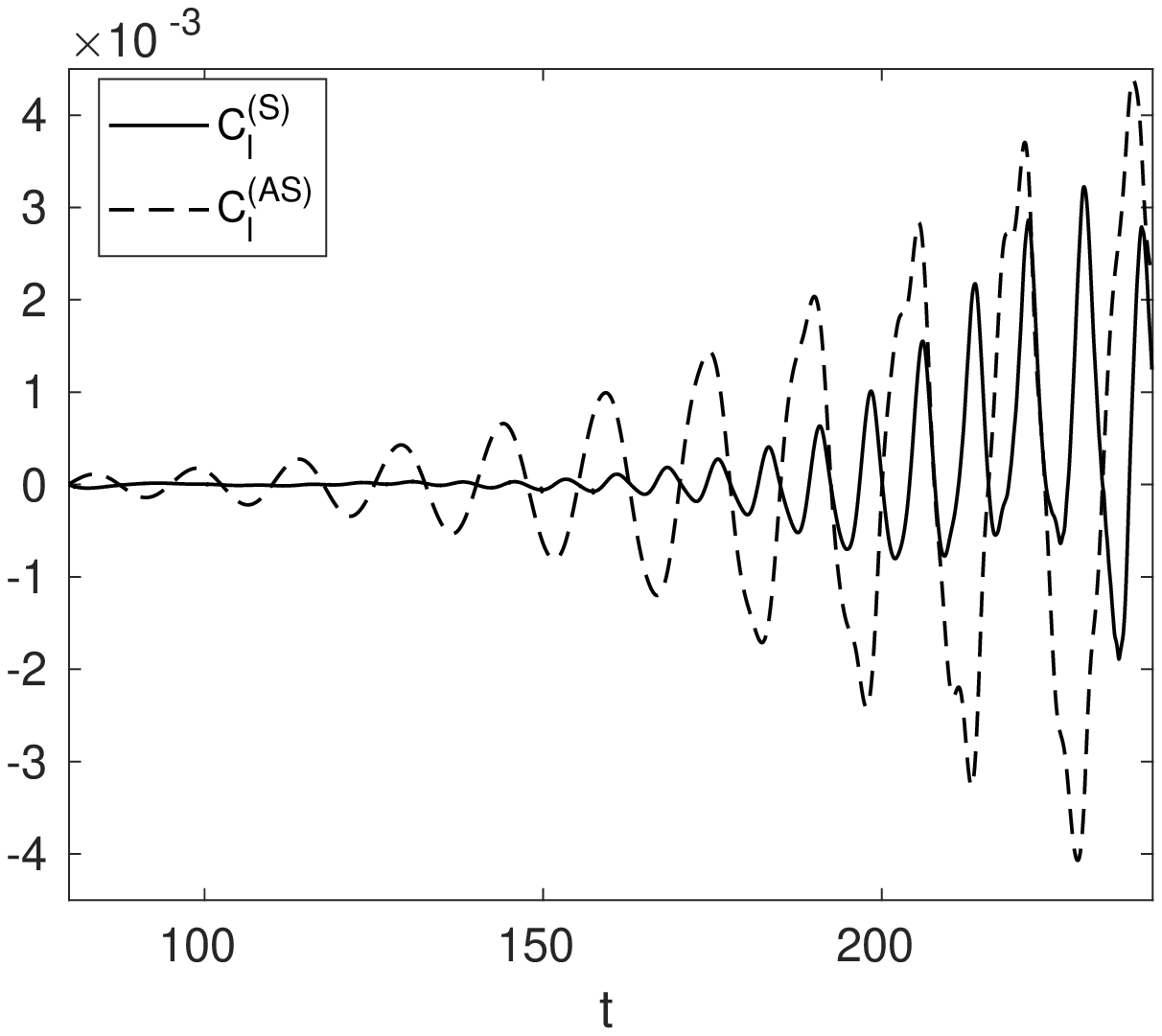}}\\
\end{tabular}
\caption{$Re=800$. Left: left and right drag (\emph{resp.} lift) coefficients $C_d^{(L)}(t)$ (\emph{resp.} $C_l^{(L)}(t)$) (black) and $C_d^{(R)}(t)$  (\emph{resp.} $C_l^{(R)}(t)$). Right: symmetric (\emph{resp.} antisymmetric) part of the drag and lift coefficients.
respectively calculated along the right ($z>0$) and left ($z>0$) parts of the BOP. 
The slowly evolving part of the lift coefficient $C_l^{\rm base}$ has been eliminated to evaluate symmetric and antisymmetric parts of the oscillations. $C_l^{\rm base}$ has been obtained by interpolation from points where the left and right part were equal.
\label{fig:halfcd800}}
\end{figure}
\begin{figure}
\begin{tabular}{cc}
\parbox{0.49\textwidth}{
\psfrag{t}{$t$}
\includegraphics[width=0.49\textwidth]{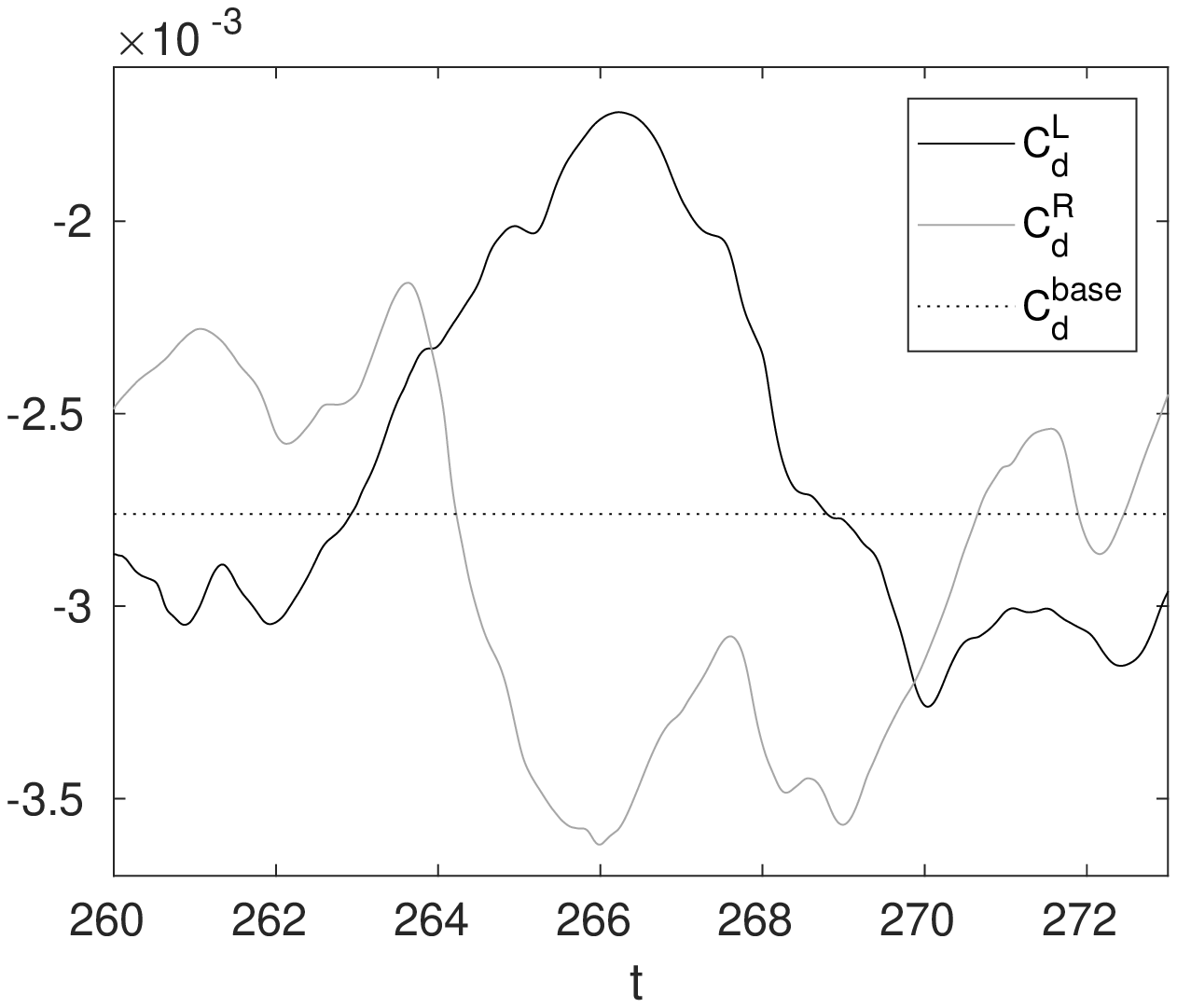}}&
\parbox{0.49\textwidth}{
\psfrag{t}{$t$}
\includegraphics[width=0.49\textwidth]{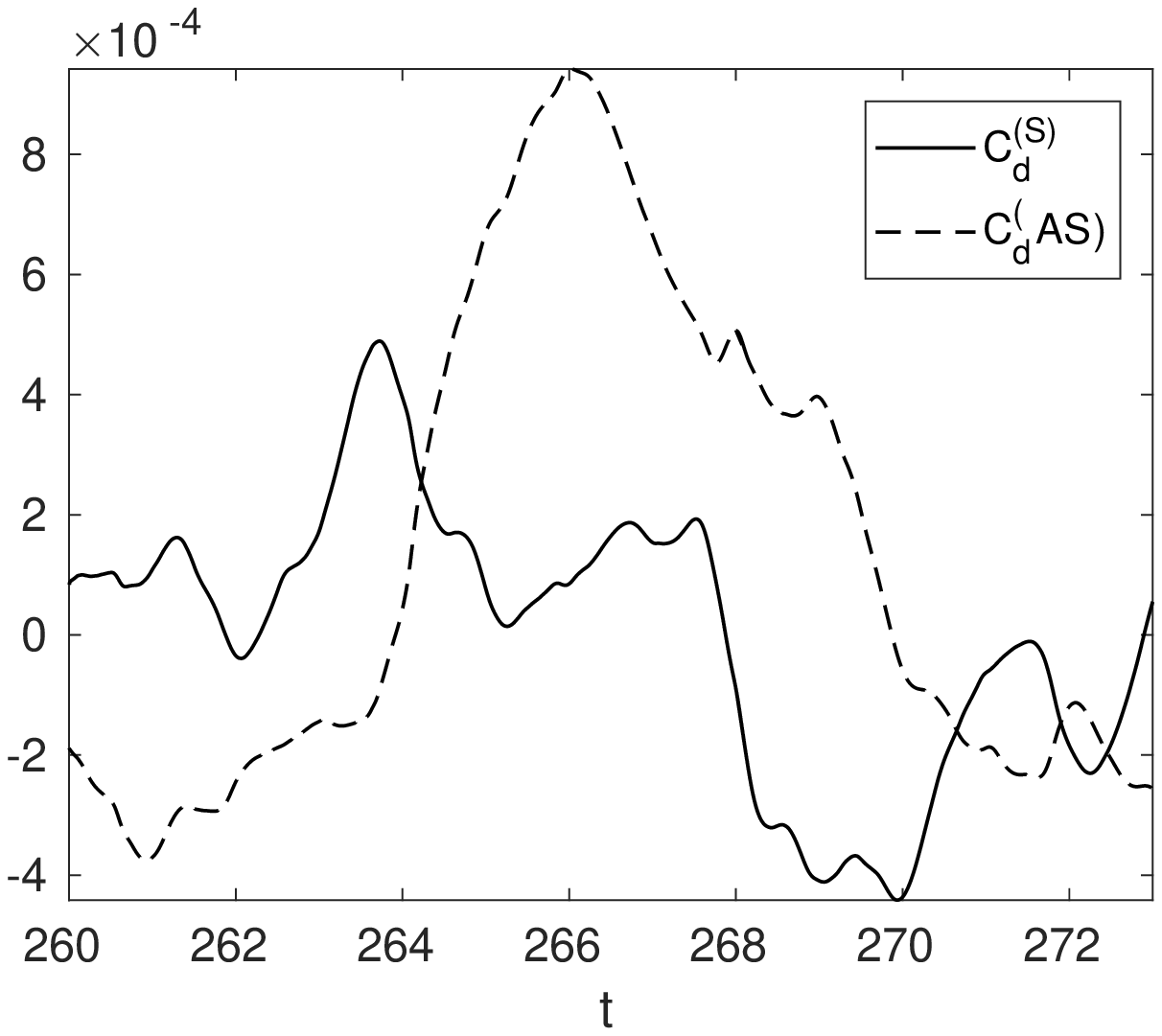}}\\
\parbox{0.49\textwidth}{
\psfrag{t}{$t$}
\includegraphics[width=0.49\textwidth]{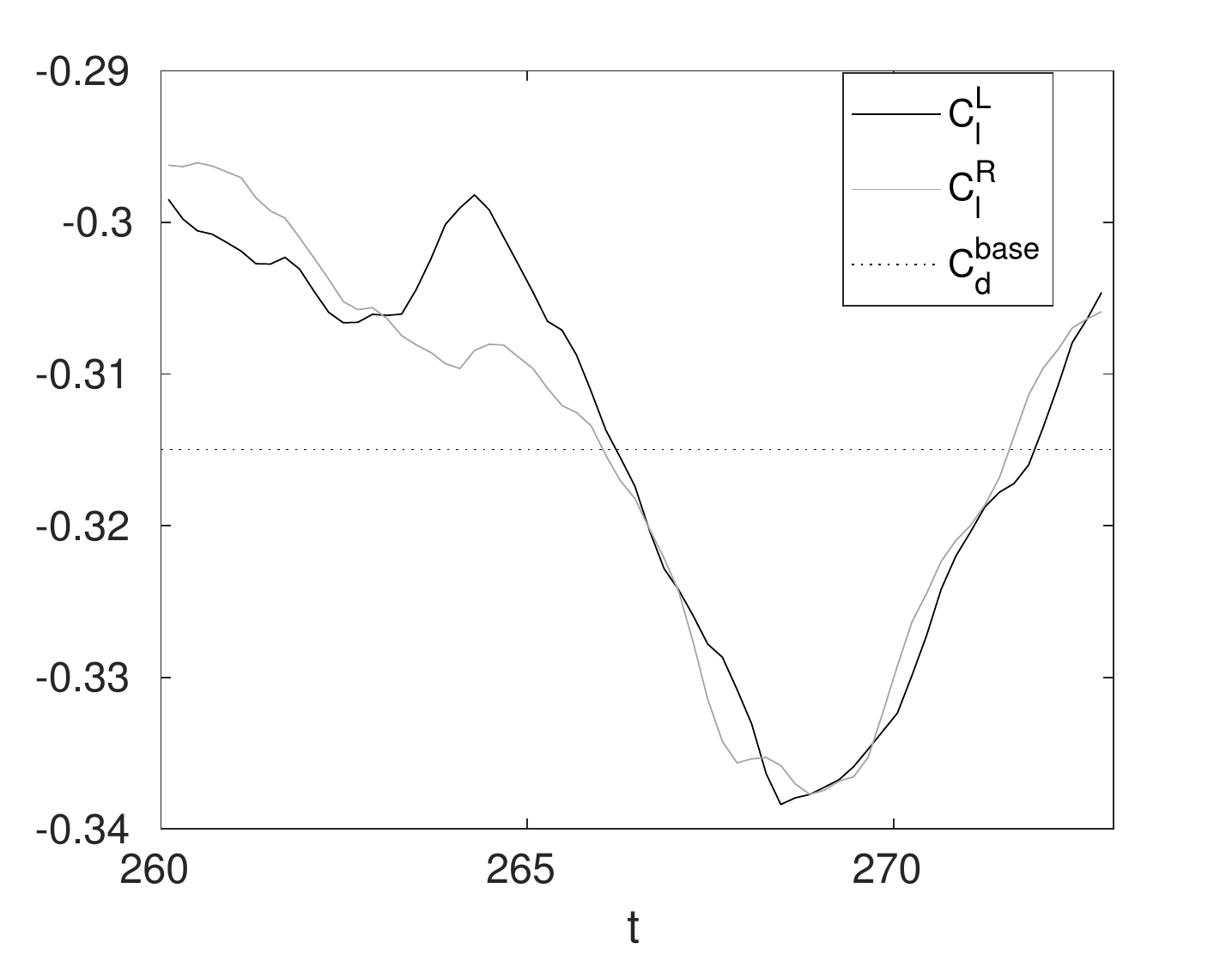}}&
\parbox{0.49\textwidth}{
\psfrag{t}{$t$}
\includegraphics[width=0.49\textwidth]{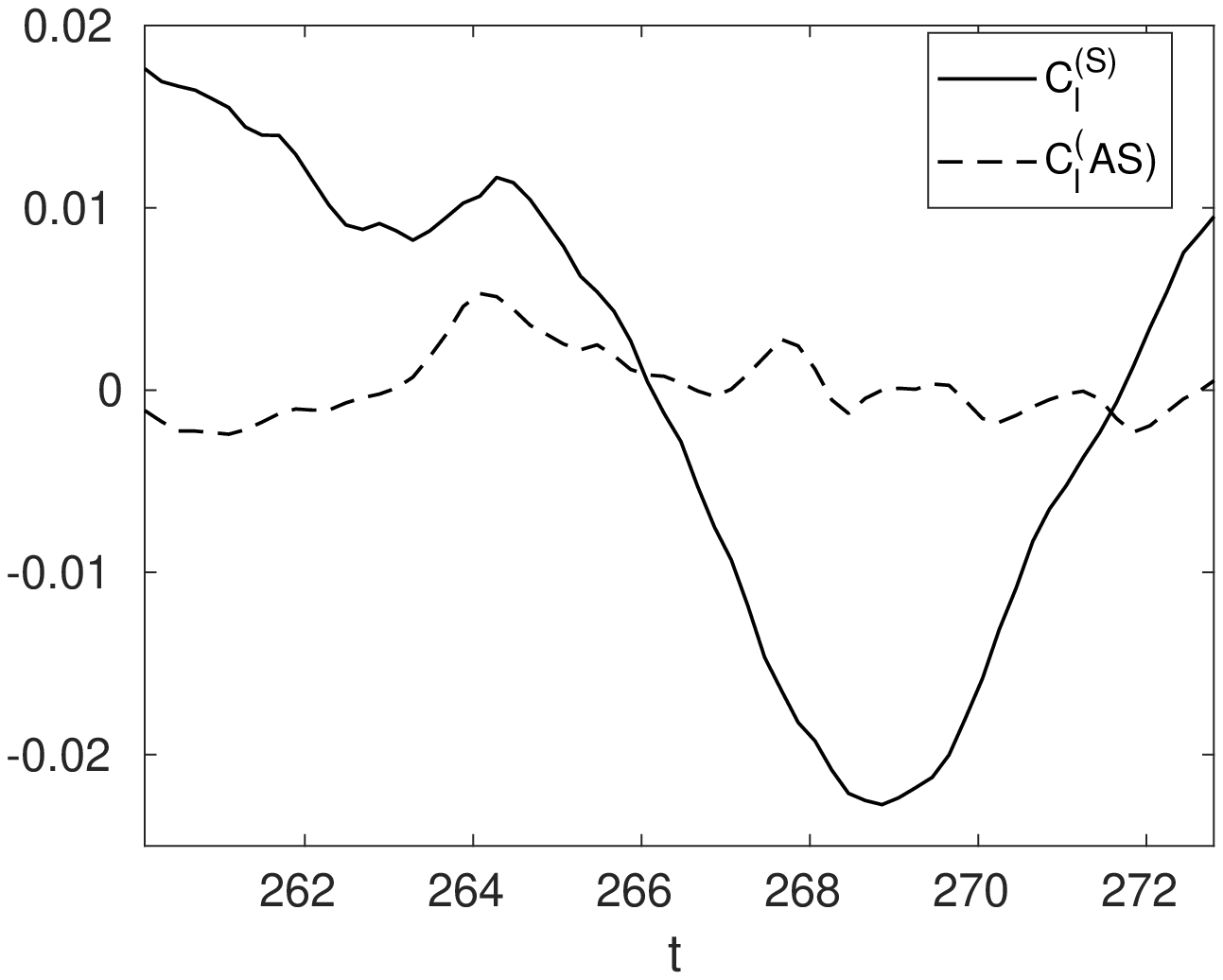}}\\
\end{tabular}
\caption{Left: left and right drag (\emph{resp.} lift) coefficients $C_d^{(L)}(t)$ (\emph{resp.} $C_l^{(L)}(t)$) (black) and $C_d^{(R)}(t)$  (\emph{resp.} $C_l^{(R)}(t)$). Right: symmetric (\emph{resp.}) antisymmetric parts of the drag and lift coefficients. $Re=2000$.
respectively calculated along the right ($z>0$) and left ($z>0$) parts of the BOP. 
\label{fig:halfcd2000}}
\end{figure}
The symmetry with respect to the CP of the oscillations is illustrated on figure \ref{fig:halfcd800}. Here, 
we have extracted the contributions to the oscillating parts of the lift and drag coefficients originating in the $z>0$ and $z<0$ halves of the BOP, respectrively $C^{L}_{d,l}$ and $C^{R}_{d,l}$. The half-sum and 
half-differences of these quantities $C^{(S)}_{d,l}$ and $C^{(AS)}_{d,l}$ respectively represent the
 symmetric and the antisymmetric parts of the drag and lift coefficients on the BOP with respect to the CP. 
The evolution of these quantities clearly shows that the oscillating part of the drag is perfectly 
antisymmetric at the onset and remains so up to around $t\simeq150$. Past this point, the exponential 
growth saturates as non-linearities become important, and the non-antisymmetric part grows until 
$t\simeq300$, when the single oscillation breaks up.
At this point, the drag, and therefore the flow have mostly lost their antisymmetric structure.\\

Because of the Dean vortices, the base steady state is very different to 
cases with no walls or periodic boundaries, either in sharp bends or in related 
problems (Backward facing step etc...), where the base state is invariant 
along z.
Our previous DNS of the flow in a sharp bend with lateral periodic boundary 
conditions (LTBC) instead of walls showed that unsteadiness appeared through a 
$z-$periodic deformation of the recirculation bubble, which soon became 
unstable to small scale three-dimensional instabilities, leading up to 
turbulence (\cite{zp2013_pf}). The presence of the side walls is stabilising
in the sense that for an opening ratio of 1, unsteadiness appears at a Reynolds 
number between $700$ and $800$ with them and, around $397$ without them (see linear 
stability analysis in \cite{shps2016_jfm}).
Nevertheless, the onset of unsteadiness in both geometries share important features:
(i) In both cases, unsteadiness occurs in the main recirculation bubble,
(ii) Both unsteady modes are antisymmetric with respect to the CP. 
In the infinite geometry, however, the spanwise wavenumber of the unstable mode is 2 for LTBC (\emph{vs} approximately unity here), and the unstable mode is not oscillatory.
%
In the end, the important differences between the phenomenologies at the onset reflect dissimilar 
instability mechanisms. This is consitent with the prominent role played by the Dean vortices in the bend of square cross-section. Indeed, even though in both cases an intrinsic instability of the recirculation bubble plays a lead role in the onset of unsteadiness, the Dean flow profoundly reshapes this region in the end of square section, and ultimately drives the onset of the instability itself.

\subsection{Vortex shedding}
\begin{figure}
\centering
\includegraphics[width=0.49\textwidth]{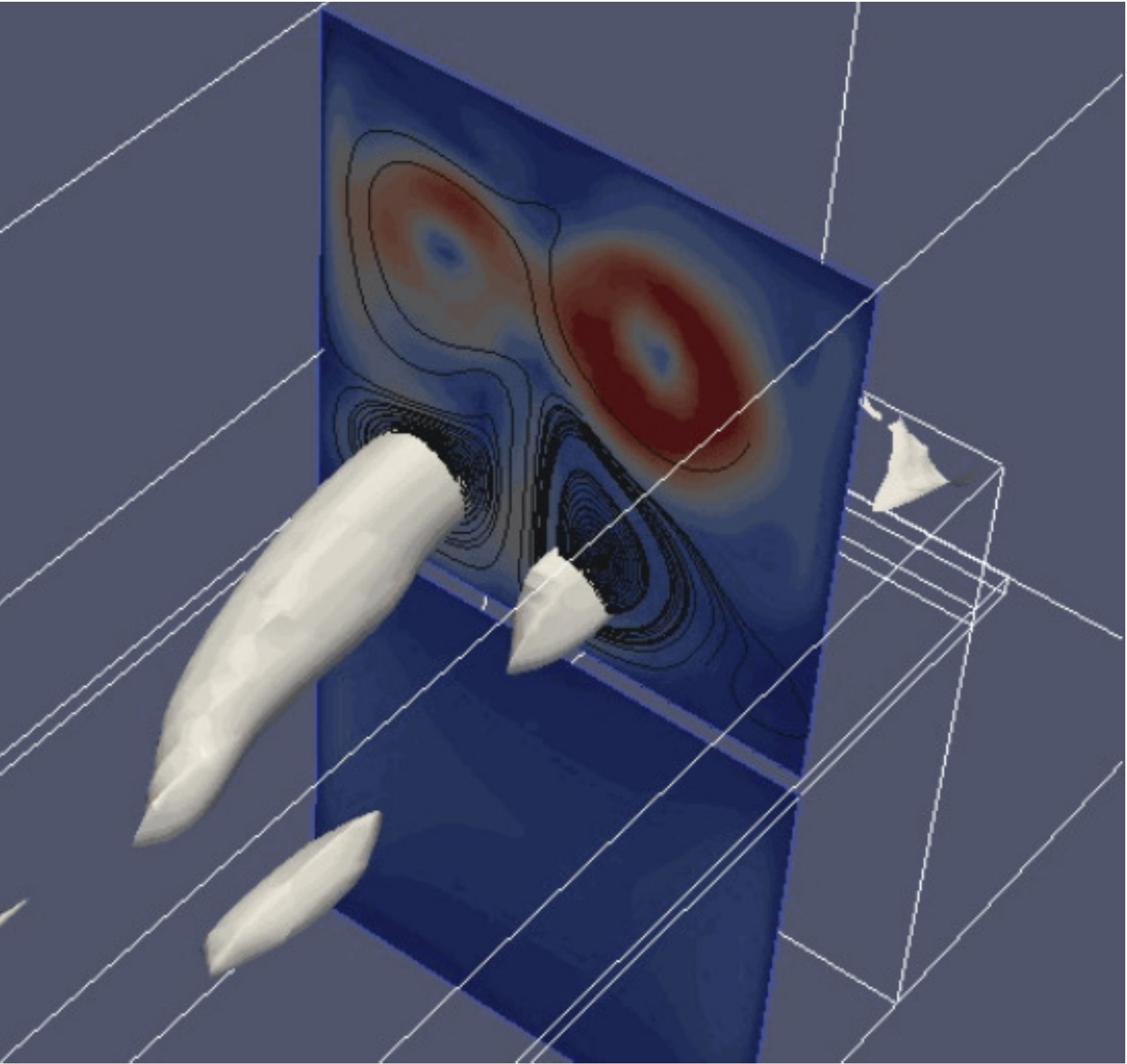}
\includegraphics[width=0.49\textwidth]{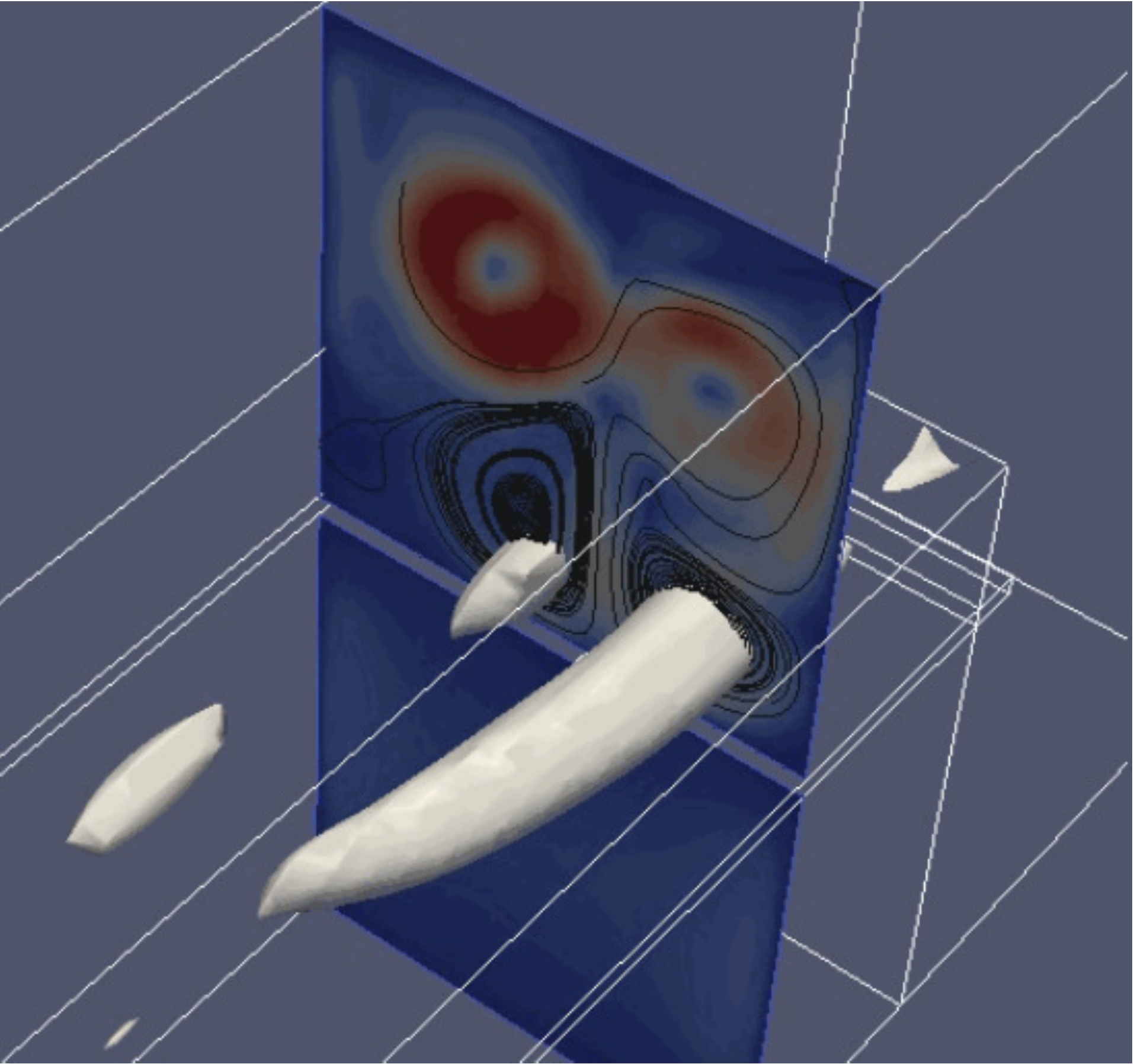}
\includegraphics[width=0.49\textwidth]{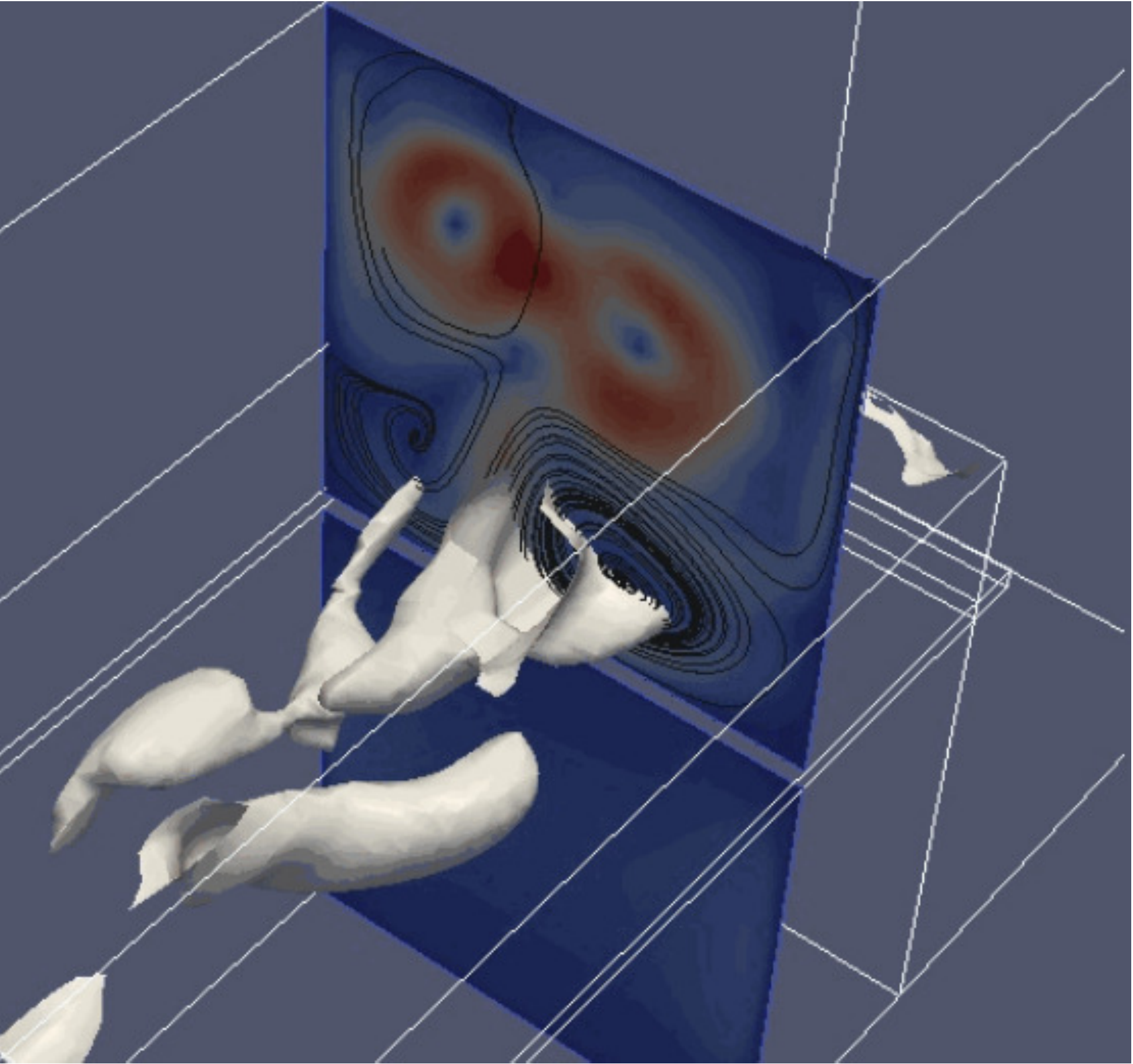}
\psfrag{0}{$0$}
\psfrag{025}{$0.25$}
\psfrag{05}{$0.5$}
\psfrag{065}{$0.646$}
\includegraphics[width=0.08\textwidth]{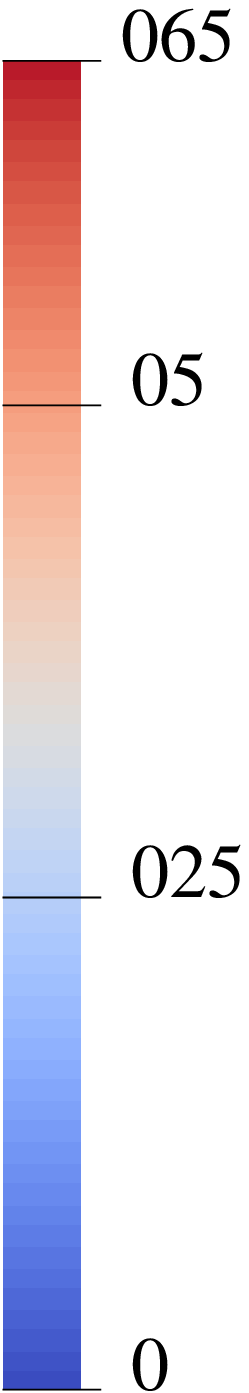}
\caption{Contours of $\lambda_2=1.5\times10^{-2}$ in grey showing three stages of the the vortex shedding process. Top left: formation of a streamwise vortex in the left lobe (at time $t=380.56$), top right: formation of a streamwise vortex in the right lobe ($t=392.16$), bottom intermediate stage ($t=377.36$). Streamlines of the 2D velocity projection in a $(y,z)$ plane show that during the formation of a streamwise vortex, both the corresponding lobe of the Dean vortex pair and the vortex on the side of the vortex being formed progressively grow in size at the expense of the other side. The switch-over between vortex formation on the left and and the right lobe is initiated as the formed vortex breaks up and sheds from the formation region (bottom picture). Colours represent streamwise velocity magnitude. 
An animation of the vortex shedding process is available as supplementary material in file "lambda2Re800shedding.avi".
\label{fig:shedding_lambda2}}
\end{figure}
Let us now focus on the oscillatory flow that follows the breakup of the harmonic oscillations at $t>300$. The time variations of iso$-\lambda_2$ surfaces on figure \ref{fig:shedding_lambda2} (see also 
associated movie in the supplementary material), soon reveal that the brutal change of drag and lift 
coefficients between $t=300$ and $t=320$ corresponds to the point where the streamwise oscillations of 
the right and left lobes of the main recirculation have become sufficiently strong to trigger their 
 breakup and subsequent shedding. The settled oscillations that ensue at $t>320$, result from the  
successive re-formation and shedding of these structures, in a similar fashion to the shedding mechanism 
in the von K\`arman street observed in the wake of a cylindrical obstacle (\cite{williamson1996_arfm}). As in this famous example, vortices are alternatively formed on the right and the left side of the
centreplane, and shedding on one side takes place while a new structure is formed on the other side.
 However, while the structures forming the von K\`arman street are spanwise (\emph{i.e} normal to the 
flow direction), the original feature of the shedding mechanism is that the structures alternately 
forming and shedding in the sharp bend 
are mostly streamwise. 
This feature is again due to the presence of the Dean flow: the formation of the streamwise vortices 
is indeed fed by the streamwise vorticity generated by the vertical jet it induces at the location 
of the first recirculation in the CP.\\
The detailed mechanisms governing the formation and shedding of these streamwise vortices are 
also more complex than those governing the Von K\`arman street. While the periodic shedding in the sharp 
bend seems to exist on a narrow but high range of Reynolds numbers from around 800 (see section 
\ref{sec:higherre}), the von K\`arman street survives in a range between $Re=46$ and $Re\simeq150$ where 
it becomes unstable to three-dimensional A and B modes. Over this range, the von K\`arman street 
essentially induces sinusoidal variations of the drag and lift coefficients, when the 
corresponding variations of $C_d$ and $C_l$ exhibit a significantly more intricate waveform for the sharp 
bend (figure \ref{fig:cd}). In a way, this makes it even more remarkable that such a clearly periodic 
shedding mechanism exists at the onset of unsteadiness in the flow in the sharp bend. 
%
%
\subsection{Flow at higher Reynolds number \label{sec:higherre}}
The well-ordered, periodic flow at $Re=800$ does not survive a moderate increase in Reynolds number. At 
$Re=1000$, periodicity is lost and the flow becomes chaotic even though the frequency spectra of lift 
and drag coefficients are sill dominated by a small number of frequencies characteristic of the shedding 
mechanism. At $Re=2000$, no such dominance stands out: the broad-band continuous aspect of the 
spectra, and the erratic fluctuations of flow coefficients are indicative of a state closer to fully 
developed turbulence (see figure \ref{fig:cd}). 
The full analysis of these regimes would take us well beyond the scope of the present work, nevertheless, it is interesting to notice that the symmetry properties that characterise the onset of unsteadiness are also mostly lost in these regime (see figure \ref{fig:halfcd2000}). This suggests that these erratic fluctuations and the turbulence that ensure may be driven by altogether different mechanisms than those governing vortex shedding. 
%
%
%
\subsection{Characterisation of the bifurcation to unsteadiness}
\begin{figure}
\centering
\includegraphics[width=0.8\textwidth]{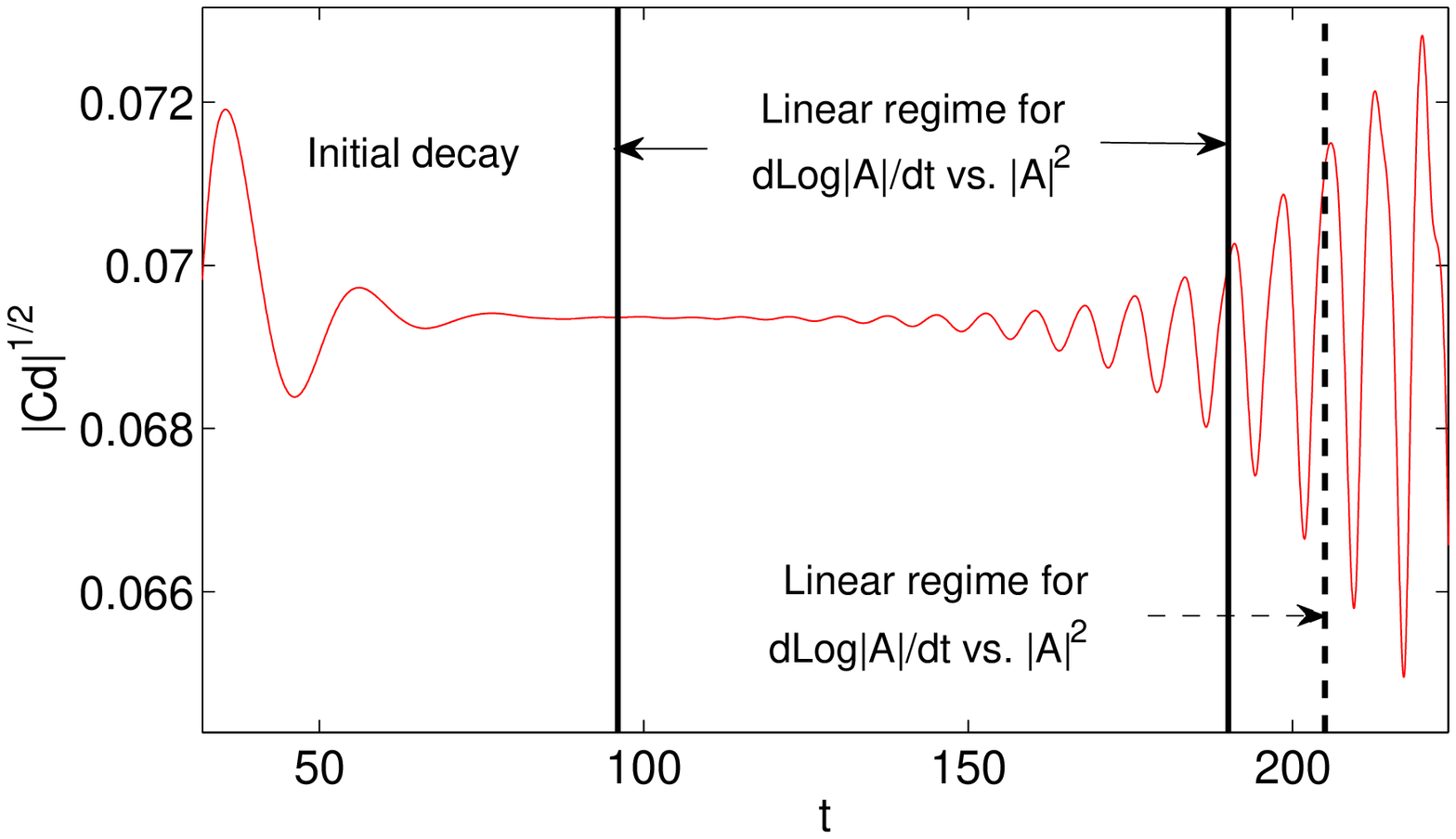}
\includegraphics[width=0.6\textwidth]{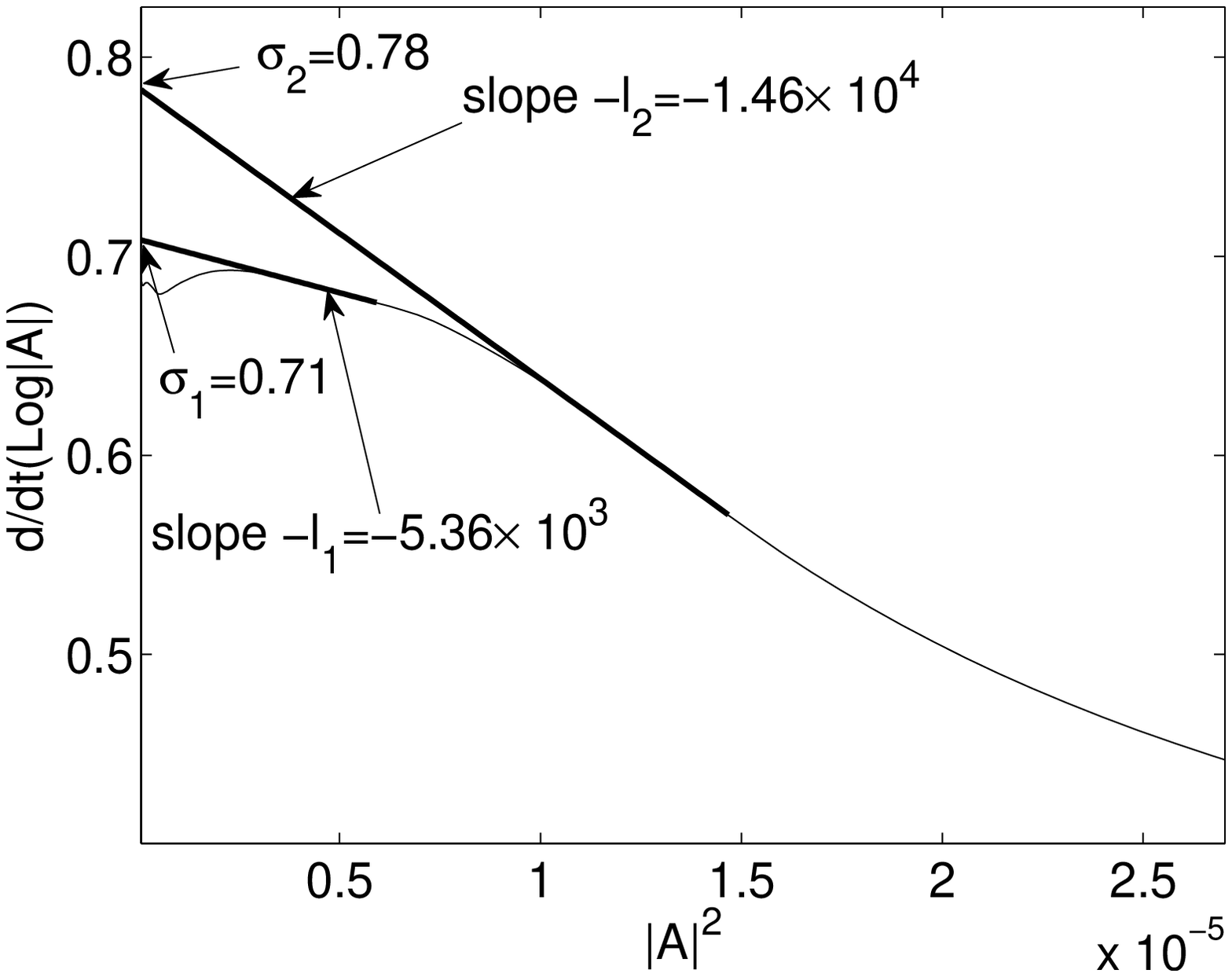}
\caption{Top: time variations of $|Cd|^{1/2}$ used for Stuart-Landau analysis, showing the two time intervals were $d(\log|A|)/dt$ varies linearly with $|A|^2$ (solid line for the firt bifurcation and dashed line for the second). Bottom: variations of $d(\log|A|)/dt$ \emph{vs.} $|A|^2$ and extrapolation to $|A|=0$ from which coefficients $\sigma$ and $l$ are obtained. \label{fig:landau}}
\end{figure}

To conclude the analysis of the unsteady flow, we shall come back to the onset of unsteadiness and seek to characterise the nature of the bifurcation leading to it. Following \cite{sheard2004_jfm}, the sub- or super- critical nature of the bifurcation is obtained by fitting the time evolution of a the
 complex amplitude $A$ of a perturbation around equilibrium, to a Landau 
equation of the form
\begin{equation}
\frac{dA}{dt}=(\sigma+i\omega) A-l(1+ic)|A|^2A+\mathcal O(A^5),
\label{eq:landau}
\end{equation}
where $\sigma$ represents the exponential growthrate of the perturbation, 
$\omega$ its base frequency while $l$ reflects the level of non-linear saturation 
and $c$ is a real constant. For $l>0$, the bifurcation is
supercritical and saturation occurs through the cubic term in (\ref{eq:landau}). If $l<0$ 
on the other hand, higher order terms are needed to saturate 
the growth and the bifurcation is subcritical. The real part of Eq. (\ref{eq:landau}) 
readily implies that 
\begin{eqnarray}
\sigma&=&\lim_{|A|^2\rightarrow0} \frac{d}{dt} \log |A|,\\
l&=&-\lim_{|A|^2\rightarrow0}  \frac{d}{d(|A|^2)}\frac{d}{dt} \log |A|.
\end{eqnarray}
Several choices are possible for the quantity whose amplitude $A$ is modelled 
in (\ref{eq:landau}) (\cite{sheard2004_jfm}). For the purpose of our analysis, we 
shall use quantity $|C_d|^{1/2}$ on the grounds that it is easily extracted 
from $Cd(t)$ and that it effectively reflects changes of flow regimes and 
unsteadiness. Classically, the values of 
$\sigma$ and $l$ are sought by studying the growth or the decay of small, 
artificially added perturbations in near-critical regimes. 
In these conditions, the high order corrections neglected in 
(\ref{eq:landau}) are indeed small, so that $d/dt( \log |A|)$ varies linearly 
with $|A|^2$. 
Here, since the critical value of $Re$ for the onset of unsteadiness is not 
known, this linear dependence may only exist over short periods of time, which 
we shall capture during the early transient of our first unsteady 
case, \emph{i.e.} $Re=800$. $|A|$ is derived from the envelope of 
$|C_d(t)|^{1/2}$, represented on figure \ref{fig:landau} (top). The close 
vicinity of $|A|=0$ cannot be reliably calculated because 1) of its very 
high sensitivity to errors on the one hand and 2) because in our calculation, 
a very small residual oscillation remains from the decay of the impulse from 
$Re=600$ to $Re=800$ (marked as "initial decay" on figure \ref{fig:landau} (top)). Extrapolating the near-linear region closest to $|A|=0$ of the graph $d/dt( \log |A|)$ \emph{vs.} $|A|^2$ to $|A|^2=0$ yields $\sigma=\sigma_1=0.71\pm0.02$ and
 $l=l_1=5.36\pm0.2\times 10^3$ figure \ref{fig:landau} (bottom). The positive 
value of $\sigma$ reflects the unstable nature of the flow and the precision 
on the value of $l$ is sufficient to conclude that $l>0$. This establishes the 
supercritical nature of the bifurcation leading to the onset of unsteadiness.\\
Interestingly, a second linear region appears in 
the graph of $d/dt( \log |A|)$ \emph{vs.} $|A|^2$. It coincides with the 
loss of antisymmetry in the oscillations in the range  $Re=150-200$ that precedes 
the onset of vortex shedding at $t\geq300$. The existence of the second linear region suggests that the variations of $A$ over this interval are dominated by this second, non-antisymmetric mode. Linearly 
extrapolating this region to $|A|^2=0$ yields a growthrate and saturation coefficients of $\sigma=\sigma_2=0.78\pm0.02$ and $l=l_2=1.46\pm0.2\times 10^4$. These values suggest that at $Re=800$, the first mode is unstable and that the flow undergoes a second supercritical bifurcation leading to the emergence of the second mode. The breakup of the oscillations itself at $t\geq300$ takes place over barely more than one oscillation, and does not lend itself to this sort of analysis. 
\section{Flow coefficients \label{sec:coefs}}
\begin{figure}
\psfrag{Re}{$Re$}
\psfrag{Cd}{$C_d$}
\includegraphics[width=0.5\textwidth]{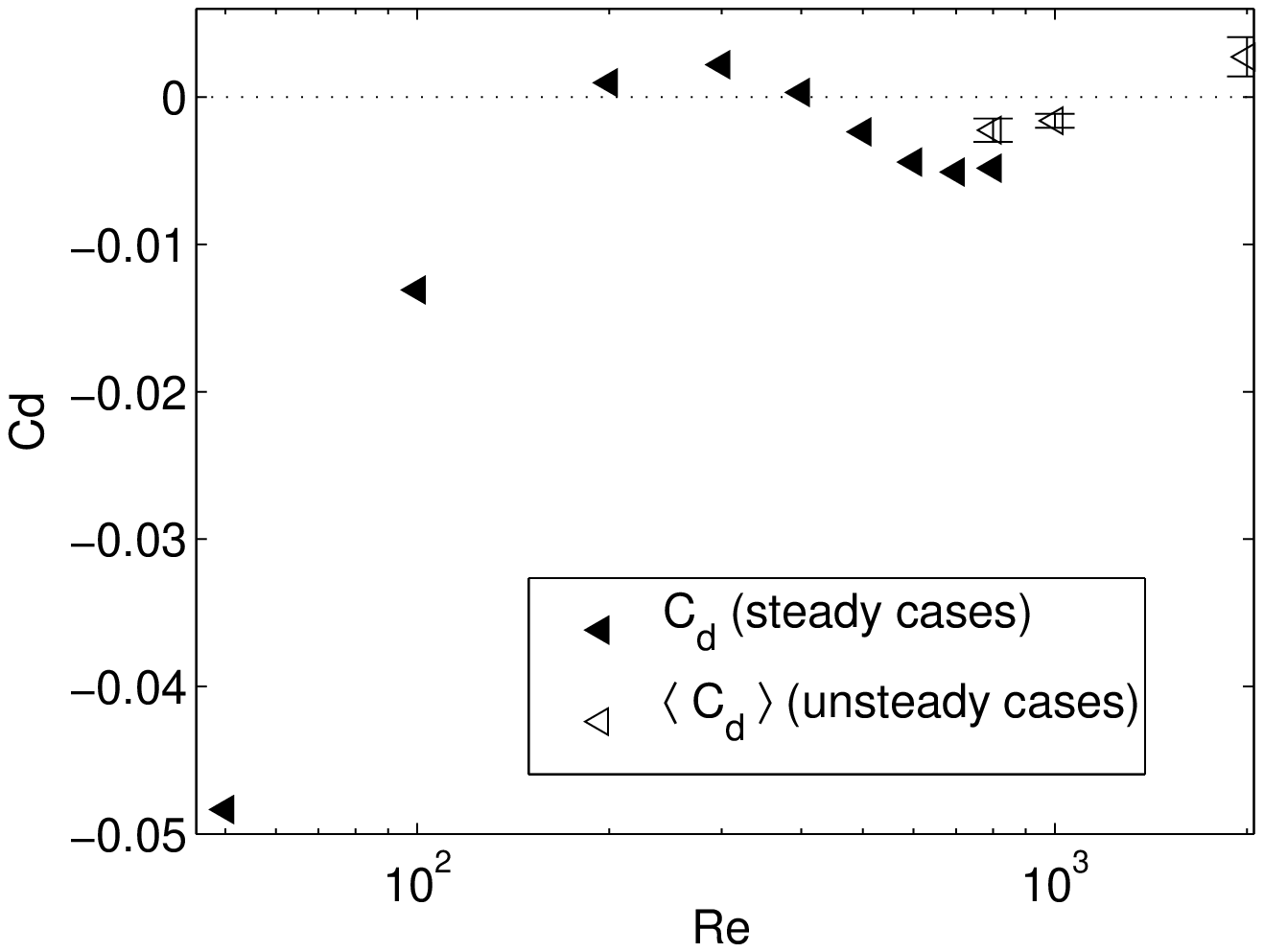}
\psfrag{Re}{$Re$}
\psfrag{Cl}{$C_l$}
\includegraphics[width=0.5\textwidth]{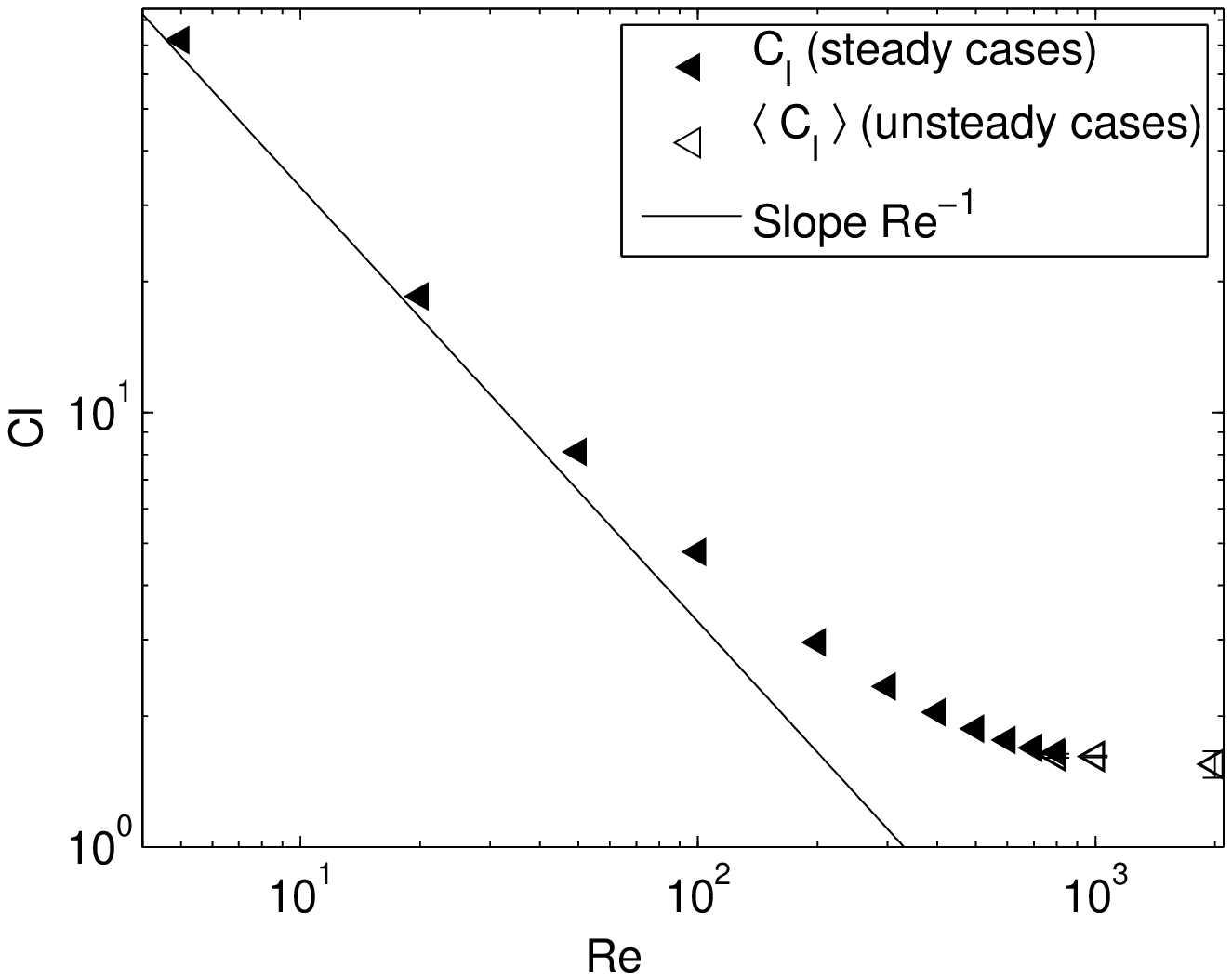}
\caption{\label{fig:cdcl} Drag (left) and lift (right) coefficients calculated on total surface of the separating element. Error bars indicate the amplitude of fluctuations of $C_d(t)$ and $C_l(t)$ in the unsteady regime, after the transient part (calculated as twice the standard deviation of these quantities).}
\end{figure}

We shall now examine how the flow phenomenology identified in the previous 
sections reflects on the classical flow coefficients used to characterise 
the different regimes of separated flows (see for example \cite{zdravkovich97_t1}).
By analogy with flows around obstacles, we shall consider the drag and lift coefficients $C_d$ and $C_l$ associated with forces on the upper surface of the bottom outlet plane (BOP). 
%
The variations of time-average 
of these quantities 
with $Re$ are represented on figure \ref{fig:cdcl}. 
At the lowest Reynolds numbers, where inertial effects are absent, the flow 
topology is independent of $Re$. The definitions of $C_d$ and $C_l$ 
imply that both quantities should scale as $Re^{-1}$ and this is indeed the
case for $Re<50$. For $Re\geq50$, inertia reshapes the flow and the first 
recirculation appears on the BOP.  Its presence mostly affects the drag coefficient: in the vicinity of the BOP, the recirculation creates a region where 
friction is in the opposite direction to the main stream and therefore reduces the overall drag on the BOP. 
As the recirculation grows in size (as measured by the distance $h_5$ between 
$SN_5$ and the leading edge of the BOP on figure \ref{fig:hwithre}),  
the drag due to the reverse flow grows, to the point of reversing the direction of the overall drag for $Re\gtrsim 200$. Similarly, the drastic shortening of the recirculation bubble at $Re\gtrsim 300$ leads to another change in sign 
of $C_d$, whose value subsequently stagnates when $h_5$ does 
(for $500 \geq Re \geq 800$, within the steady regime). 
At the onset of unsteadiness, $h_5$ increases again and the overall drag 
increases towards positive values, to become positive between $Re=1000$ and $Re=2000$. In all calculated cases, the fluctuations of  $C_d$ were found to 
remain smaller than its mean value (Respectively $35\%$, $30\%$ and $48\%$ for $Re=800,$ 1000, and 2000).\\
The variations of $C_l$ are, by contrast, practically not affected by the 
complex dynamics of the recirculation bubble and reflect mostly the progressive transition between a creeping and inertial flow (for which $p\sim \rho U_0^2$ and hence $C_l$ becomes independent of $Re$). $C_l$ is hardly sensitive 
to the unsteadiness of the flow too, with a fluctuating part representing 
$1\%$, $0.4\%$ and $7\%$ of its mean for $Re=800,$ 1000, and 2000. 

\section{Conclusion}
We have conducted a detailed analysis of the steady flow structure and the onset of unsteadiness in a 
180$^o$ sharp bend of square cross section. Besides the numerous applications of this generic 
configuration, its fundamental interest lies in the co-existence of of two classical phenomena of 
fluid dynamics: on the one hand, a recirculating bubble exist in the outlet where the flow separates 
from the inner edge of the bend. On the other hand, the strong curvature of the streamlines drives a 
so-called \emph{Dean Flow} in the turning part. This structure is made of a pair of counter-rotating 
streamwise vortices that extends into the outlet where it interacts with the recirculating bubble.\\
The main point to retain from this study is that both the steady regime and the onset of unsteadiness are entirely determined by this interaction. 
In the steady regime, a critical point analysis revealed that the complex topology of the streamlines in the 
180$^o$ sharp bend was almost entirely described by three pairs of critical points, each made of a focus and a half-saddle in the symmetry plane of the bend (CP). The first of these pairs to appear, in the sense of increasing Reynolds numbers is located near the bottom outlet plane is associated to the recirculation bubble 
($Re\simeq50$).  It is followed by a similar pair located near the bottom corner of the turning part 
($Re\geq100$), for which the associated focus is the location where the two vortex tubes forming the 
Dean vortex pair meet in the symmetry plane. A third pair of focus-half-saddle was found at $Re\geq300$,  in the upper corner of the turning part. This pair of critical points  generates a pair of vortices of 
somewhat similar topology to the Dean vortices, which we have named \emph{Bullhorn} in reference to their shape. Nevertheless these remained of much lower intensity than the Dean vortices and never reached 
sufficiently far into the outlet to exert any significant influence on the flow dynamics. 
By contrast, the flow rate associated to the Dean vortices reaches up to a third of the inflow and the 
vertical jet driven by the counter-rotating pair in the outlet soon becomes sufficiently strong to split 
the recirculating bubble into two symmetric lobes (This answers questions (i) and (ii) set in introduction).\\
The flow structure shaped by the static interaction between the Dean vortices and the recirculating 
bubble was also found to be crucial for the onset of unsteadiness as the latter first originates 
in the periodic and anti-symmetric streamwise oscillation of these lobes. Though initially sinusoidal in shape, this oscillation is soon subject to a secondary instability that breaks antisymmetry and eventually leads to the 
break-up and shedding of the lobes. Stuart-Landau analysis reveals that both the onset of unsteadiness 
and the destabilisation of the oscillating flow occur through supercritical bifurcations. 
 It remains,
however, unclear whether the breakup itself directly results from this secondary instability. 

The end result of this process is a periodic street of
 streamwise vortices that are alternately formed and shed on the left and right sides of 
the outlet. Though reminiscent of the von K\`arman vortex street, this particular vortex street 
is made of streamwise vortices whose formation is driven by the vertical flow induced by the 
pair of Dean vortices located above the recirculation bubble. In this sense, this original vortex 
formation and shedding mechanism is driven by the Dean flow, even though the onset of instability originates in the first recirculation bubble, as in infinitely extended sharp bends, where Dean flows are absent. This answers question (iii) set in introduction.

Simulations at slightly higher Reynolds numbers suggest that this mechanism is active over a rather narrow range of Reynolds numbers around $Re=800$, and that the flow at higher Reynolds numbers is turbulent, 
with fluctuations driven by different mechanisms. The trace of these regimes was found to be well 
reflected in the evolution of the drag coefficient along the outlet bottom plane with $Re$ (question (iv)). The questions that remain are those of the exact conditions in which this remarkable periodic 
shedding occurs: just how wide is the range of Reynolds number where it remains stable ? Further, since it is absent in bends that are infinitely extended in the spanwise direction, how small an aspect ratio of 
the duct section is indeed required for this mechanism to be observed ? Similarly, while large opening ratios are not expected to obstruct its dynamics (because for large values, the turning part of the flow concentrates in a region of opening 
ratio slightly smaller than 1), it is not clear whether the periodic shedding survives at arbitrary small opening ratios.\\

AP acknlowledges support from the Royal Society, through the Wolfson Research Merit Award scheme (Grant Ref WM140032).



\end{document}